\documentclass[twocolumn,times]{autart}
\usepackage[T1]{fontenc}
\usepackage[utf8]{inputenc}
\usepackage[british]{babel}
\usepackage{amsmath,amsfonts,amssymb}
\usepackage{xtab}
\usepackage{array}
\usepackage{comment}
\usepackage{mathtools}
\usepackage{times}
\usepackage{xcolor}
\usepackage{graphicx}
\usepackage{bm}
\usepackage{url}
\usepackage{xspace}
\usepackage{soul}
\usepackage[square,numbers,compress]{natbib}
\makeatletter
\def\bibfont\scriptsize
\def\@bibliosize{\scriptsize}
\makeatother
\renewcommand{\bibfont}{\scriptsize}
\usepackage[inline]{enumitem}
\usepackage{todonotes}
\newcommand{\covid}{COVID-19}
\newcommand{\sarscov}{SARS-CoV-2}

\def\N{\mathbb{N}}

\def\Pr#1{\ensuremath{\mathrm{Pr}\left \{#1 \right \}}}

\def\Card#1{\ensuremath{\left\vert #1 \right\vert}}

\newcommand{\Prc}[2]{\ensuremath{\mathrm{Pr}\left \{#1 \!\left. \right\rvert\! #2 \right \}}}

\renewcommand{\vec}[1]{{\mathbf #1 }}

\newenvironment{proof}{\textbf{Proof}:}{\hfill$\square$}

\newtheorem{theorem}{Theorem}

\newcommand{\dan}[1]{\color{black}{{#1}\xspace}\color{black}\xspace}
\newcommand{\lui}[1]{\color{black}{{#1}\xspace}\color{black}\xspace}

\newcommand{\mf}[1]{\color{black}{{#1}}\color{black}\xspace}

\newcommand{\mr}[1]{\color{black}{{#1}}\color{black}\xspace}

\DeclareMathOperator{\LGlobally}{\mathbf{G}}
\DeclareMathOperator{\LEventually}{\mathbf{F}}

\DeclareMathOperator{\PProb}{\mathbf{P}}

\usepackage{etoolbox}
\newcommand{\zerodisplayskips}{%
\setlength{\abovedisplayskip}{3pt}%
\setlength{\belowdisplayskip}{3pt}%
\setlength{\abovedisplayshortskip}{3pt}%
\setlength{\belowdisplayshortskip}{3pt}%
}
\appto{\normalsize}{\zerodisplayskips}
\appto{\small}{\zerodisplayskips}
\appto{\footnotesize}{\zerodisplayskips}

\allowdisplaybreaks[4]

\begin{document}

\begin{frontmatter}

\title{
  A Markovian Model for the Spread of the SARS-CoV-2 Virus
  \thanksref{footnoteinfo}} 

\thanks[footnoteinfo]{%
  Corresponding author L. Palopoli. Tel. +39-0461-283967.  Fax
  +39-0461-282093.}

\author[UniBZ]{Luigi Palopoli}\ead{luigi.palopoli@unitn.it},
\author[UniTN]{Daniele Fontanelli}\ead{daniele.fontanelli@unitn.it},
\author[UniBZ]{Marco Frego}\ead{marco.frego@unibz.it},
\author[UniTN]{Marco Roveri}\ead{marco.roveri@unitn.it}

\address[UniTN]{University of Trento, Via Sommarive 9, 38122 Povo (TN), Italy}
\address[UniBZ]{Free University of Bolzano, NOI Techpark, via Volta 13, 39100 Bolzano (BZ)}


\begin{abstract}                          
  We propose a Markovian stochastic approach to model the spread of a
  \sarscov{}-like infection within a closed group of humans. The model
  takes the form of a Partially Observable Markov Decision Process (POMDP),
  whose states are given by the number of subjects in different
  health conditions. The model also exposes the different parameters
  that have an impact on the spread of the disease and the
  various decision variables that can be used to control it (e.g,
  social distancing, number of tests administered to single out
  infected subjects).
  \lui{
  The model describes the stochastic phenomena that underlie the spread of the
  epidemic and captures, in the form of deterministic parameters,
  some fundamental limitations in the availability of resources (hospital beds and test swabs).
  The model lends itself to different uses. For a given control policy,
  it is possible to \emph{verify} if it satisfies an analytical property
  on the stochastic evolution of the state (e.g., to compute probability that the hospital beds will reach
  a fill level, or that a specified percentage of the population will die). If the control policy is not
  given, it is possible to apply POMDP techniques to identify an optimal control policy
  that fulfils some specified probabilistic goals.
  Whilst the paper primarily aims at the model description, we show with numeric examples
  some of its potential applications.
}

\end{abstract}

\end{frontmatter}

\section{Introduction}
The catastrophic outbreak of the \covid{} pandemic caused by the
\sarscov{} virus has recently brought to the attention of academia,
decisions makers, and common people the importance of reliable models
to predict and control the evolution of the disease.  Far from being
an academic exercise, such models are key to define policies that
reduce the risk of spreading the virus, minimising the impact on the
economy and on the quality of social life.
To be up to the task, a model has to meet a few requirements.
First, it shall capture the multi-faceted nature of human
relations. Most of human interactions take place within a limited
circle of people (family members, colleagues, friends), and yet our
lives are punctuated by random encounters and accidental events. A
realistic model has to account both for the regularity and for the
elements of randomness of human relations.
Second, the level of details shall be sufficient to expose the impact
of all the different parameters affecting the spread of the virus
(e.g., use of masks, number of intensive care beds), and of the
commands (e.g., social distance, number of tests administered) used to
control its evolution.
Finally, the model shall be analytically tractable for
the synthesis of control policies and the evaluation of
various scenarios.

\noindent
\textbf{Paper Contribution. } In this paper, we propose a stochastic model for the
evolution of a \covid{}-like epidemic. The model describes the
evolution of a population of subjects, each one allowed to be in eight
different states.  Such parameters, as the probability of contracting
the infection during a single meeting with asymptomatic subjects
(which is directly connected to the use of protective masks) or the
number of available beds in intensive care facilities, are first class
citizens in the model.  Besides, the model exposes two of the decision
variables usable to control the evolution of the disease: social
interaction limits and number of tests. Since part of the states are
directly observable and part are not, the model takes the form of a
partially observable Markov Decision Problem (POMDP).
\lui{Our model is control--oriented, in that it exposes the
  impact of the command decisions on the transition probabilities, and
  hence on the evolution of the epidemic. Its main areas of
  applications are two-fold. First, given a control policy (e.g., a
  decision to apply lockdown depending on the estimated number of
  people affected by the disease), it is possible to exactly compute the
  probabilities and properties of interest,
  ranging from the probability that the deceased individuals will be
  beyond an acceptable threshold, to more general properties expressed
  in propositional temporal logic~\cite{browne1988characterizing}.
  Second, it is possible to synthesise control policies where such
  properties are respected by construction~\cite{ahmadi2020control}.
  Both goals can be fulfilled by using PRISM, STORM or similar tools.\\
  Such methods are based on the study of the analytical evolution of the
  POMDP, therefore they are guaranteed to provide probabilistically
  correct results, insofar as the model is sufficiently close to the
  reality.  Our primary effort in this paper has been toward modelling
  with a sufficient level of details such effects as the transition
  between the different states and the resource limitations (e.g.,
  hospital beds or daily availability of tests).  A potential price to
  pay for a detailed model is its level of mathematical tractability.
}
Indeed, our model remains tractable for groups having a population in
the range of a few tens of people, which is the typical size of a
circle
of people that have daily interactions.  \\
\lui{The study of larger populations using the presented model is
  possible in different ways. A first possibility, discussed in the
  related work section, is by exploiting some recent results on
  scalable techniques for MDP analysis and
  design~\cite{nasir2020epidemics,Hak18}.  Another possibility is by
  using a small size population model to develop a policy with
  guaranteed properties, and then recover the result on a heuristic
  basis for larger populations verifying its efficacy through Monte
  Carlo simulations, as shown in this paper}.


\noindent
\textbf{Related work}
The first pioneering work on mathematical epidemiology dates back to
\citeauthor{bernoulli:1760}~\cite{bernoulli:1760} with his study of
smallpox models. However, most of the theory was developed only in the
last century with the introduction of compartmental
models~\cite{Kermack:1927}.  \lui{The idea of a compartmental model is
  to break down a population into several compartments.  The simplest
  is the Susceptible-Infected-Recovered (SIR) model, used under the
  assumption that a recovered subject is not infected again (in the
  short-term), while the Susceptible-Infected-Susceptible (SIS) model
  is used when a recovered subject can be infected again.  The
  migration of subjects between the different compartments is ruled by
  a set of Differential Algebraic Equations (DAE), that is a set of
  ODEs connected by an algebraic relation (the sum of the cardinality
  of the compartments yields the population).  The outbreak of the
  SARS epidemic in 2002-2003 sparked a new interest in these
  models~\cite{allen:2008,Brauer:2019}, with the main goal of the
  researchers being a strategy to keep in check the evolution of the
  epidemic leveraging the control variables.  } The use of
deterministic models paves the way for a number of interesting control
theoretical results such as models with many
compartments~\cite{giordano:2020}, flow
control~\cite{ghezzi:1997,giordano:2016} and
stability~\cite{Khanafer:2016}, policies for \covid{} based on Optimal
Control~\cite{Yousefpour:2020}.
These models have an important role in the identification of key
epidemiological features that indicate the progress of the disease,
such as the basic reproduction number. However, they are only a rough
approximation of the reality because random factors naturally
influence the evolution of an epidemic~\cite{anderson:1992}.
While deterministic models represent the mean disease dynamics, a much
richer set of information can be derived from stochastic models.
%
As an example, stochastic models may converge to a disease-free state even if the
corresponding deterministic models converge to an endemic equilibrium~\cite{anderson:1992}.
%
%
In addition, stochastic models produce a variety of important information,
such as the probability of an outbreak, the distribution of the final
size of a population or the expected duration of an
epidemic~\cite{Brauer:2019,Sattenspiel:1990}.
Extensions of these models have been developed to account for stochastic
fluctuations in some of the parameters~\cite{Ji:2012,Liu:2012}, but even
so, these models do not entirely capture the natural randomness inherent in the very
way humans interact.\\
\lui{
  The \sarscov{}/\covid{} pandemic has given another boost to the
  research in the area, highlighting the importance of statistical
  techniques to elaborate the massive amount of data daily collected
  in most of the countries affected by the pandemics, in order to
  support the decision making process and the definition of long-term
  policies.  A notable example is the work
  of~\citeauthor{ZARDINI2021100530}~\cite{ZARDINI2021100530}, where
  the authors apply statistical methods to quantify the probability of
  transition between different state of \covid{}-affected patients based
  on the age class. Another interesting example is a statistical study
  to evaluate the effects of lockdown policies in Italy after the
  outbreak of the pandemic~\cite{Flavia20}.
  Although these methods use compartmental models, they cannot directly be
  used for control design or to verify closed--loop properties.\\
}
\lui{An important class of models amenable to analytical analysis are
  the ones that respect the so-called Markov property, which we
  generally define Markov
  Processes~\cite{cassandras2009introduction,allen:2008}. When we observe the
  system in discrete--time, Markov Models are called
  discrete-time Markov chains (DTMC). In continuous--time, we talk
  about continuous--time Markov chain (CTMC). Contrary to other stochastic models
  such as Stochastic Differential Equations (SDE),
  Markov models (DTMC and CTMC) adopt a numerable state space
  composed of discrete variables. When command variables become part
  of the model, Markov chains morph into Markov Decision Processes
  (MDP). Since we observe the system at periodic intervals,
  the modelling technique adopted in this paper is based on discrete--time Markov Decision Processes and,
  since not all states are directly observable (e.g., asymptomatic persons), we have a POMDP.\\  }
There are, historically, two paradigms for modelling a disease spread as
a DTMC, the Reed-Frost model and the Greenwood model. They are called
chain-binomial models because the transition probabilities are
governed by binomial random variables. In these models, infectious
($I$) are indexed in generations capable of infecting susceptibles
($S$) for one generation after which they are not involved in the
process. If the population has constant size $N$, then the initial
condition is $S_0+I_0=N$ and $S_{t+1}+I_{t+1}=S_t$, for times
$t=0,1,\ldots$. Hence $S_t+\sum_{\tau=0}^t I_\tau=N$. The number of
infectious at time $t+1$ is given by a binomial random variable of
parameters $S_t$ and $p(I_t)$ (the probability that a susceptible
becomes infected if the infectious are $I_t$). The corresponding
transition probability is given by the $k$-th component of the
Binomial probability mass function (pmf):
\begin{equation}
  \label{eq:BinomialPMF}
  \begin{aligned}\textstyle
  & \Prc{I_{t+1}=k}{S_t=x,I_t=y} =  \\
  & \textstyle \quad \mathcal{B}(x,p(y))_k = \binom{x}{k}p(y)^k(1-p(y))^{x-k} ,
  \end{aligned}
\end{equation}
which is trivially zero if $k > x$. A difference between the Greenwood
and the Reed-Frost models is that in the former $p(y)\equiv p$ is
constant, while in the latter we have that $p(y)=1-(1-p(1))^{y}$, where $p(1)$ is
the probability that a susceptible is infected by one infective.
Extensions of this model were presented
by~\cite{gani:1971,Tuckwell:2007}. \lui{As discussed in Section~\ref{sec:markov_tested}, the
model adopted in this paper can be seen as a generalisation of the Reed-Frost model.\\}
The use of stochastic models opens for the possibility to use
Stochastic Model Checking in order to study probabilistic temporal
properties to evaluate the effects of a strategy on a population
during the evolution of a disease. As an
example,~\citeauthor{10.1371/journal.pone.0145690}~\cite{10.1371/journal.pone.0145690}
adapted a Susceptible-Exposed-Infectious-Recovered-Delayed-Quarantined
(Susceptible/Recovered) CTMC model used to analyse the spread of
internet worms using the PRISM model
checker~\cite{kwiatkowska2011prism}. The same tool was used to
validate a model and to compute the minimum number of influenza
hemagglutinin trimmers required for fusion to be between one and
eight~\cite{C1MB05060E}. Finally,
\citeauthor{Chauhan2015}~\cite{Chauhan2015} uses a stochastic simulation
to compute timing parameters for a timed automaton, which is then
verified against some temporal properties.\\
All the stochastic models mentioned are abstract and simplified.  For
instance, the use of three states (Susceptible, Infectious, Recovered)
certainly makes the model tractable,
but does not provide enough details on the parameters and the command
variables influencing the evolution of the disease.  Another
simplification frequently adopted to make the model tractable by model
checkers is to enforce that only one subject can change her/his state
across one transition (which is unrealistic), or to avoid the
possibility of defining command variables.  \lui{Only recently have
  more sophisticated models emerged.  An interesting example is the
  work of~\citeauthor{Viet18}~\cite{Viet18}, in which the authors use
  an MDP to model the spread of the Porcine Reproductive and
  Respiratory Syndrome (PPRS) and use it to synthesise a regional
  policy for its containment.  A similar attempt has been made also
  for the \sarscov{}
  by~\citeauthor{nasir2020epidemics}~\cite{nasir2020epidemics}. The
  authors consider a segmented model for the population and model its evolution
  via a MDP with the purpose of synthesising an optimal policy for vaccination, hospitalisation
  and quarantine. The model is largely simplified and the paper omits any discussion on the
  computation of the transition probability as a function of the policy, which is essential
  for the definition of the MDP.
  Our work lies in this line of research. As in~\cite{nasir2020epidemics}, we show how to
  construct a MDP describing the \sarscov{} epidemics using a segmented model for the population,
  but our model is very detailed and contains an exhaustive discussion on how to compute the transition
  probabilities and on how to use the model for control design and analysis.
  Very interestingly,~\citeauthor{nasir2020epidemics}~\cite{nasir2020epidemics} discuss in alarming
  terms the problem of scalability of epidemic models based on MDPs. The technique that they suggest to deal with large
  populations is by abstraction, i.e.,  by considering a unit of population in
  the model as the representative of a number of people. This is essentially equivalent to
  a process of quantisation in which the accumulated error can be controlled and shown to
  be tolerable, for practical purposes. Clearly, the same line of reasoning potentially applies
  to the model presented in this paper.\\
  Another interesting technique to address the problem of scalability has been
  recently suggested by~\citeauthor{Hak18}~\cite{Hak18} and is based on Graph Based MDPs.
  The authors consider a network of similar MDPs (i.e., MDP governed by the same update rules)
  that interact with their neighbours following a graph topology. By exploiting symmetry and
  the graph topologies, the authors show that the design of a sub-optimal policy for the MDPs
  can be achieved by approximate linear programming. This technique could be applied in our
  context considering the model proposed in this paper as the nodes of the graph.
}

\noindent
\textbf{Paper Organisation}
The paper is organised as follows. After introducing notation and
definitions in Section~\ref{sec:problem}, we describe the proposed stochastic
model in Section~\ref{sec:markov_tested}, with the transition probabilities and
computational remarks on the complexity of the model.  The model checking techniques
for the problem at hand and some numeric results to illustrate the potential
application of the  proposed model are reported in
Section~\ref{sec:numeric}. Finally, Section~\ref{sec:Conclusion}
presents the discussion about the model and describes the lines of
future work. In Appendix~\ref{sec:appA} there is the complete proof of the main theorem.
In Appendix~\ref{sec:markov_untestes}, related to the computational complexity,
we propose also a simplified model that may be used for computational
efficiency.

\section{Definitions and Notations}
\label{sec:problem}

We model a population as a stochastic discrete--time system with a
finite state.  The evolution is observed at discrete time $k$ and each
subject can belong to one of eight possible states.  We take into
account the usual classes of susceptible, infected, recovered ($SIR$),
the asymptomatic subjects $A$ (i.e., a group of infected people that
do not exhibit symptoms but are infective), hospitalised $O$ and dead
$D$ people and we further add the group of people that recover from an
asymptomatic state $Ra$ and the case of swab-tested people, that are
quarantined ($Q$) if they result positive.  The subjects who are in a
state at step $k$ will be denoted by a calligraphic letter (e.g.,
$\mathcal{S}_k$ is the set of susceptible subjects).
The table in Figure~\ref{fig:sym} reports the symbols
used to denote the different sets, their cardinality (e.g., $S_k$ is the cardinality of $\mathcal{S}_k$) and the different probabilities governing the transition of a subject
between the different sets.
\begin{figure}[tb]
\bgroup
\newcolumntype{C}{>{$}l<{$}}  
\newcolumntype{L}{@{}l}
\setlength\tabcolsep{5pt}
\renewcommand{\arraystretch}{0.98}
\scalebox{0.815}{
\fbox{\begin{tabular}{CC@{}}
  \multicolumn{2}{L}{\textbf{Sets}}\\
   S_k  = \Card{\mathcal{S}_k}  & \text{N. of susceptible sub. $\mathcal{S}_k$ at step } k,\\
   A_k  = \Card{\mathcal{A}_k}  & \text{N. of asymptomatic sub. $\mathcal{A}_k$ at step } k,\\
   I_k  = \Card{\mathcal{I}_k}  & \text{N. of symptomatic sub. $\mathcal{I}_k$ at step } k,\\
   R_k  = \Card{\mathcal{R}_k}  & \text{N. of recovered sub. $\mathcal{R}_k$ at step } k,\\
   Ra_k = \Card{\mathcal{R}a_k} & \text{N. of asympt. recovered sub. $\mathcal{R}a_k$ at step } k,\\
   O_k  = \Card{\mathcal{O}_k}  & \text{N. of hospitalised sub. $\mathcal{O}_k$ at step } k,\\
   D_k  = \Card{\mathcal{D}_k}  & \text{N. of deceased sub. $\mathcal{D}_k$ at step } k,\\
   Q_k  = \Card{\mathcal{Q}_k}  & \text{N. of quarantined sub. $\mathcal{Q}_k$ at step } k,\\
   Q^{(R)}_k \!\!=\! \Card{\mathcal{Q}^{(R)}_k\!}\! & \text{N. of quarantined sub. recovered $\mathcal{Q}^{(R)}_k$ at step } k.\\
   \multicolumn{2}{L}{\textbf{Deterministic Parameters}}\\
   N & \text{Total number of subjects},\\
   C & \text{Available beds in hospital facilities}.\\
    \multicolumn{2}{L}{\textbf{Probabilistic Parameters}}\\
   \omega &\text{Prob. to contract the infection in one meeting},\\
   \beta &\text{Prob. for an infectious asympt. sub. to recover},\\
   \delta &\text{Prob. for an asympt. sub. to devel symptoms},\\
   \mu &\text{Prob. for a symptomatic sub. to recover},\\
   \alpha &\text{Prob. for a symptomatic sub. to die},\\
   \sigma &\text{Prob. for an hospitalised sub. to die},\\
   \xi &\text{Prob. for an hospitalised sub. to recover},\\
   \gamma &\text{Prob. for a tested infectious sub. to be positive},\\
   \psi &\text{Prob. for a symptomatic sub. to be hospitalised},\\
   \iota & \text{Prob. that a quarantined sub. devel symptoms},\\
   \upsilon &\text{Prob. that a quarantined sub. recovers}.\\
   \multicolumn{2}{L}{\textbf{Command Variables}}\\
   M & \text{Num. of people met by any subject\textsuperscript{*}},\\
   t & \text{Num. of people tested\textsuperscript{*}}.
\end{tabular}
}}
\egroup
\caption{Summary of symbols. \textsuperscript{*}Can be variable with the state.}
\label{fig:sym}
\end{figure}
Notice that, in order for the systems to be consistently defined,
the following constraints shall hold on the probabilities:
$C_{\beta, \delta} = (1 -\beta - \delta) \ge 0$,
$C_{\mu, \psi, \alpha} = (1 -\mu - \psi -\alpha) \ge 0$,
$C_{\sigma,\xi} = (1-\sigma-\xi) \ge 0$, and
$C_{\iota,\upsilon} = (1-\iota-\upsilon) \ge 0$. These constraints will be used in the rest of the paper to prove the theorems.
In our model two deterministic variables $M$ and $t$ are used as
control variables, which represent respectively the number of meetings
allowed in each period and the number of tests administered.  \lui{In
  addition, two deterministic parameters ($N$, $C$) are demographic
  parameters related to the size of the population and to the
  availability of hospital beds. The other parameters are
  probabilities. For instance $\omega$ represents the probability for
  a subject to contract the infection in one meeting. Such
  probabilities depend on many factors (e.g., use of masks,
  vaccination) and can be estimated by means of statistics, such as
  the one presented
  by~\citeauthor{ZARDINI2021100530}~\cite{ZARDINI2021100530}.\\}
In the rest of the paper, we consider that the binomial coefficient
$\binom{a}{b}$ for generic $a \geq 0$ and $b \geq 0$ is: equal to $0$
when $b > a$; equal to $1$ when $a = b = 0$.  Moreover, we recall the
definition of the following quantities: a) the dispositions counting
the different ways of arranging $k$ items from a set of $n$ elements
are denoted with
$\mathbb{D}_{n,k} = n(n-1) \dots (n-k+1) = \frac{n!}{(n-k)!}$; b) the
permutations of $n$ elements $\mathbb{P}_n = n!$. Therefore, the
binomial coefficient is defined as
$\binom{n}{k} = \frac{\mathbb{D}_{n,k}}{\mathbb{P}_k}$ and the
multinomial coefficient (i.e. the permutations with repetitions
obtained computing all the permutations of $n$ elements taken from $k$
sets with $n_1, n_2,\dots,n_k$ elements with
$n_k = n - \sum_{i=1}^{k-1} n_i$) is defined as
$\mathbb{M}_{n,n_1,n_2,\dots,n_{k-1}} = \binom{n}{n_1,n_2,\dots,n_k} =
\frac{\mathbb{P}_n}{\prod_{i=1}^k \mathbb{P}_{n_i}}$.

\section{Extended SAIROD Markov Model}
\label{sec:markov_tested}

\begin{figure}
  \centering
  \includegraphics[width=0.99\columnwidth]{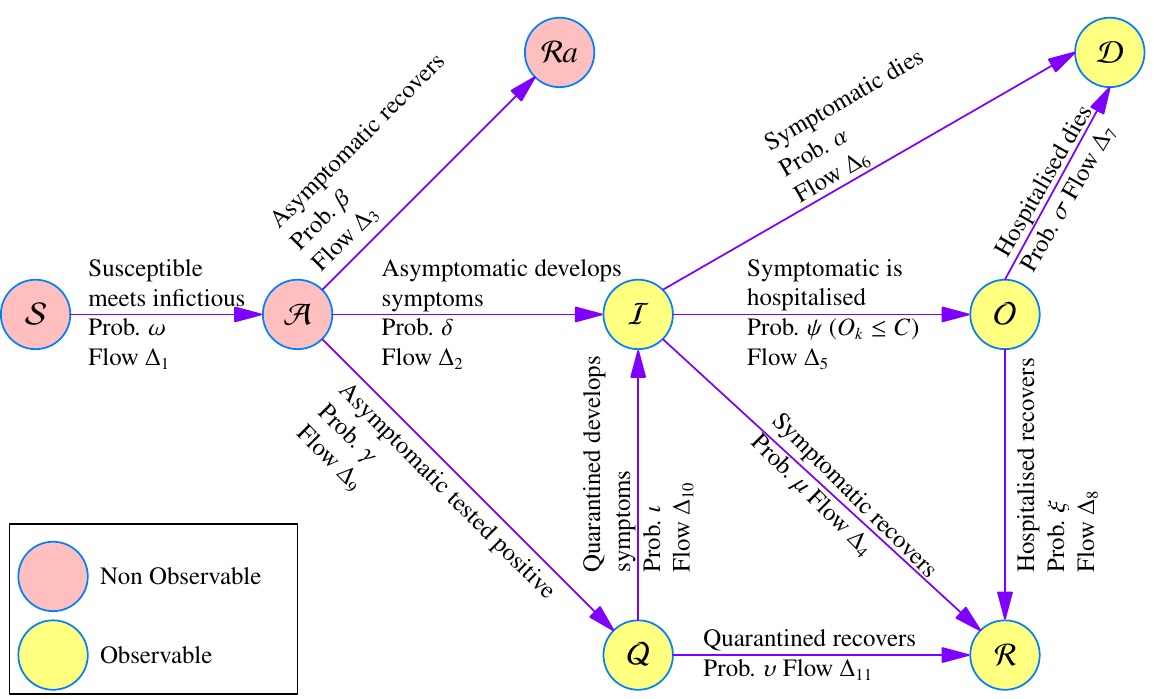}
  \caption{State transitions and flows (denoted with $\Delta_i$)
    between the different states of the Markov model.}

   \label{fig:stateTest}
\end{figure}

In this section we discuss our Markov model representing the spread of
a \covid{}-like infection. We will make the following assumptions that
reflect the best knowledge currently available on \covid{} and on its
management:
\begin{enumerate*}
\item if a tested subject is found infectious, she/he becomes
quarantined and is isolated from other people until recovery (or
death),
\item quarantined persons are constantly monitored;
\item persons
that are recovered cannot be re-infected by the virus in the short
period.
\end{enumerate*}
\lui{The first two assumptions are pretty natural and are typically
  enshrined within the regulation of most of the \covid{}-affected
  countries. The third one restricts our analysis to an interval of
  time in which virus variant forms do not materialise.}  Furthermore,
we do not consider other external death causes, and that there are no
false positives for the test outcomes.  \lui{Should this assumption be
  invalidated, the results of our analysis would be conservative in
  terms of the probability of deaths and therefore acceptable.\\}
The possible states of a subject are the ones depicted in
Figure~\ref{fig:stateTest}.
\lui{ Since we are observing the system at discrete--time and, under
  the assumptions made, the number of people in each state determines
  the evolution of the population regardless of how this
  configuration has been reached, our system can be conveniently
  modelled as a discrete--time Markov chain.}
  The state of the Markov
chain will be associated with the 8-tuple collected in the following
vector
\[
\vec{V}_k = \left[S_k, A_k, I_k, R_k, O_k, D_k, Q_k, Ra_k \right],
\]
where the values of all the different quantities are non-negative integers representing
the cardinality of their respective sets (see Figure~\ref{fig:sym}). The elements of the set
are bound to respect  the following constraints:
\begin{equation}
  \begin{aligned}
   &S_k + A_k + I_k + R_k + Ra_k+ Q_k + O_k + D_k = N,\\
    &O_k \leq C,
  \end{aligned}
  \label{eq:domain_invariantTested}
\end{equation}
that is, the population cardinality is constant, and
the number of hospitalised people is limited by a (positive)
capacity $C$.
\lui{We assume that the presence of a virus can be detected either if the subject starts to develop
  symptoms of the disease or when the subject is tested positive. The subject can be traced in all the states in which
  the presence of the virus has been diagnosed: $\mathcal{Q}$, $\mathcal{I}$, $\mathcal{O}$, $\mathcal{D}$, $\mathcal{R}$. Such states
  are \emph{observable}.
  On the contrary, since it is not possible to distinguish a subject who is susceptible,  asymptomatic or recovered without
  having developed symptoms, the states $\mathcal{S}$, $\mathcal{A}$, $\mathcal{R}_a$ are not observable.}
In the rest of this section, we show how to compute the transition
probability:
\begin{equation}
  \label{eq:transProbDef}
  \Prc{\vec{V}_{k+1}=\vec{v}'}{\vec{V}_k = \vec{v}},
\end{equation}
where $\vec{v}$ and $\vec{v}'$ are (resp.) the state vector values
before and after the transition.
%
%
Let
\[
  \vec{\Delta v} = \vec{v}'-\vec{v}\! =\!
  [ \Delta_S, \Delta_A, \Delta_I,\Delta_R,\Delta_O,\Delta_D,\Delta_Q,\Delta_{Ra}]^T
\]
and consider Figure~\ref{fig:stateTest}, where, by the notation
$\Delta_i$, we denote the flows of subjects between the different
states (e.g., $\Delta_1$ is the number of susceptible subjects who
become infected).  Each $\Delta_i$ is a number defined in the set
$D= [0,\,N] \cap \N$.  Thereby, we define a new vector
$\vec{\Delta} =
[\Delta_1,\,\ldots,\,\Delta_{11}]$ 
whose components respect the balance equations \eqref{eq:balanceTests},
labelled by $B_i$ for $i=1,\ldots, 8$ and collectively denoted as $B$:
\begin{align}
  B_1:\quad & -\Delta_1                                          &=&\quad \Delta_S, \nonumber\\
  B_2:\quad &  \Delta_1 -\Delta_2 -\Delta_3 - \Delta_9           &=&\quad \Delta_A, \nonumber\\
  B_3:\quad &  \Delta_{10}+\Delta_2 -\Delta_4 -\Delta_5 -\Delta_6 &=&\quad \Delta_I, \nonumber\\
  B_4:\quad &  \Delta_{4} +\Delta_8 +\Delta_{11}                  &=&\quad \Delta_R, \nonumber\\
  B_5:\quad &  \Delta_5 -\Delta_7 - \Delta_8                     &=&\quad \Delta_O, \label{eq:balanceTests}\\
  B_6:\quad &  \Delta_6 + \Delta_7                               &=&\quad \Delta_D, \nonumber\\
  B_7:\quad &  \Delta_9 - \Delta_{10} - \Delta_{11}               &=&\quad \Delta_Q, \nonumber\\
  B_8:\quad &  \Delta_3                                          &=&\quad \Delta_{R_a}. \nonumber
\end{align}
\noindent $B_1$ captures that the flow
from set $\mathcal{S}_k$ (susceptible) is only towards $\mathcal{A}_k$
(asymptomatic).
 $B_2$ captures that the population in $\mathcal{A}_k$ is increased by
the inflow from $\mathcal{S}_k$ and decreased by the outflow to
$\mathcal{Q}_k$, $\mathcal{I}_k$ and $\mathcal{R}a_k$, meaning that an
asymptomatic subject can either be tested positive and be quarantined
(increasing $Q_k$), or exhibit symptoms and be classified as infected
(increasing $I_k$) or recover (increasing $R_a$).
$B_3$ captures that the subjects developing symptoms ($\mathcal{I}_k$)
receives an inflow from set $\mathcal{A}_k$ and from set
$\mathcal{Q}_k$, and has an outflow toward people who develop severe
symptoms and are hospitalised (set $\mathcal{O}_k$), who recover (set
$\mathcal{R}_k$) and who die (set $\mathcal{D}_k$).
$B_4$ shows that, in our model, the set $\mathcal{R}_k$ has no outflow
and receives inflows from $\mathcal{Q}_k$, $\mathcal{I}_k$ and
$\mathcal{O}_k$.
$B_5$ shows that the set of hospitalised people has an inflow from the
infected state and produces two possible outcomes, a recovery or
death.
$B_6$ captures that the set $\mathcal{D}_k$ has no outflows, and only
two inflows, from $\mathcal{I}_k$ and from $\mathcal{O}_k$.
The set $\mathcal{Q}_k$ accounts for the quarantined subjects (i.e.,
tested positive). These subjects are constantly monitored and are not
allowed to meet until they test negative. No transitions to
quarantined state other than the ones that are asymptomatic subjects
are modelled since no false positive test outcomes are considered.
Therefore, the balance $B_7$ for $\mathcal{Q}_k$ is given by an inflow from $\mathcal{A}_k$ and two out flows towards $\mathcal{I}_k$
and $\mathcal{R}_k$.
Finally, $B_8$ concerns the set $\mathcal{R}a_k$ including subjects
that recover directly from the asymptomatic state. Therefore, they are
assumed to be immune in the short-time, but there is no possibility of knowing that this
is the case. As a consequence, contrary to the subjects who recovered
from quarantine or known illness ($\mathcal{R}_k$), these subjects may
be tested again and again with negative results. To summarise, the set
$\mathcal{R}a_k$ has no outflows and has inflows only from
$\mathcal{A}_k$. A direct consequence of this model is that the flows
from the different states are subjected to the following constraints:
\begin{equation}
  \label{eq:mr1Tests}\renewcommand{\arraystretch}{0.95}
  \begin{array}{r@{\hspace*{5mm}}r}
    \Delta_1 \le S_k,                  & \Delta_2 + \Delta_3 + \Delta_9 \le A_k,\\
    \Delta_{10} + \Delta_{11} \leq Q_k, & \Delta_4 + \Delta_5 + \Delta_6 \le I_k, \\
    \multicolumn{2}{c}{\Delta_7 + \Delta_8 \le  O_k.}
  \end{array}
\end{equation}
They ensure that the number of subjects remaining in the sets $\mathcal{S}_k$, $\mathcal{A}_k$, $\mathcal{I}_k$, $\mathcal{O}_k$ and
$\mathcal{Q}_k$ is always non-negative. These
constraints are not necessary for the absorbing sets
$\mathcal{R}_k$, $\mathcal{D}_k$ and $\mathcal{R}a_k$.

We denote by $l(\cdot)$ an assignment of variables:
$\Delta_i = \delta_i$, for $i = 1,\,\ldots,11$.  For an assignment of
variable $l(\cdot)$ we use $l(\cdot) \models \varphi$ to mean that the
assignment $l(\cdot)$ satisfies formula $\varphi$. We say
that an assignment is feasible if it satisfies the
equations~\eqref{eq:balanceTests}.
For instance,
$l\left(\Delta_1=\delta_1,\,\Delta_2=\delta_2,\,\Delta_3=\delta_3\right)
\models B_2$ means that the assignment
$\delta_1,\,\delta_2,\,\delta_3$ to the variables $\Delta_1$,
$\Delta_2$ and $\Delta_3$ satisfies balance equation $B_2$.  Likewise,
by $\vec{\Delta} \models B$ we mean that the vector
$\vec{\Delta} = [\Delta_1,\dots,\Delta_{11}]^T$ respects all the
balance equations.

It is easy to re-write the definition of the transition
probability~\eqref{eq:transProbDef} as follows:
\begin{equation}\label{eq:transProbReDef}
  \Prc{\vec{V}_{k+1}\!=\!\vec{v}'}{\vec{V}_k \!=\! \vec{v}} \!=\!
  \Prc{l(\vec{\Delta})\models B}{\vec{V}_k \!=\! \vec{v}}.\!\!\!\!
\end{equation}
For notational simplicity, we assume that $\vec{V}_k$ has been
assigned to a particular value $\vec{v}$ and use the following
notation: $\Prc{l(\vec{\Delta}) \models B}{\vec{V}_k}$ as a shorthand for~\eqref{eq:transProbDef}. Before stating and proving the theorem
that allows us to compute the transition
probability~\eqref{eq:transProbReDef}, let us introduce the following
notations:
\begin{itemize}

\item $l_1$ is an assignment linking the variable $\Delta_1$ defined
  via $B_1$ as: $l_1$: $(\delta_1 = -\Delta_S)$;

\item $l_2$ is a function linking the variables $\Delta_2$,
  $\Delta_3$ and $\Delta_9$ (with the variable $\Delta_9$ obtained
  via equation $B_2$, and the variable $\Delta_3$ obtained via
  equation $B_8$),
  defined as: \\
  $l_2$:
  $(\Delta_2=\delta_2, \Delta_3 = \Delta_{R_a},\,\,
  \Delta_9=\!-\!\Delta_S\!-\!\Delta_A \!-\!\Delta_{R_a} \!-\!
  \delta_2)$;

\item $l_3$ is a function linking $\Delta_4,\Delta_5,\Delta_6$ and
  given by: \\ $l_3$:
  $(\Delta_4=\delta_4,\Delta_5=\delta_5, \Delta_6=\delta_6)$;

\item $l_4$ is an assignment linking the remaining variables defined as:\\
  $l_4$:
  $(\Delta_7 = \Delta_D - \delta_6,\,\, \Delta_8 = \delta_5 +\delta_6
  - \Delta_D - \Delta_O,\,\,
  \Delta_{10}=\Delta_I-\delta_2+\delta_4+\delta_5+\delta_6,
  \Delta_{11}=\Delta_R + \Delta_D + \Delta_O - \delta_4 -\delta_5 -
  \delta_6 )$;
\item $l_5$, finally, assigns $(\Delta_2=\delta_2, \Delta_3=\delta_3,
      \Delta_9=\delta_9)$.

\end{itemize}

We are now able to express~\eqref{eq:transProbReDef} in a computable
form.
\begin{theorem}
  \label{teo:trasProb}
The transition probability~\eqref{eq:transProbReDef} is rewritten as
      \begin{align}
      & \Prc{l(\vec{\Delta}) \models B}{\vec{V}_k = \vec{v}} = \nonumber\\
     & \quad = \Prc{l_1}{\vec{V}_k} \cdot \displaystyle\sum_{\delta_2=0}^{-\Delta_S-\Delta_A-\Delta_{R_a}}
      \Prc{l_2}{\vec{V}_k \wedge l_1} \nonumber\\
     & \quad \cdot \displaystyle\sum_{\delta_4=0}^{\delta_2 - \Delta_I}
      \displaystyle\sum_{\delta_5=0}^{\delta_2 - \Delta_I-\delta_4} \sum_{\delta_6=0}^{\delta_2 - \Delta_I-\delta_4-\delta_5}
      \!\!\!\!\!\!  \Prc{l_3}{\vec{V}_k \wedge l_1 \wedge l_2}  \label{eq:transProbFinalExt}\\
     &  \quad \cdot \Prc{l_4}{\vec{V}_k \wedge l_1 \wedge l_2 \wedge l_3}\!\! \nonumber
    \end{align}
\end{theorem}

\begin{proof}
By applying the Total Probability Law in the conditioned case
to~\eqref{eq:transProbReDef}, we have:
\[
  \begin{aligned}
    & \Prc{\vec{V}_{k+1}=\vec{v}'}{\vec{V}_k} = \Prc{l(\vec{\Delta}) \models B}{\vec{V}_k} = \\
    & \sum_{\delta_1 \in D} \!\!\Prc{l(\vec{\Delta}) \models B}{\vec{V}_k
      \!\wedge\! (\Delta_1=\delta_1)}\Prc{\Delta_1=\delta_1
    }{\vec{V}_k}\! .
  \end{aligned}
\]
From the first balance equation~\eqref{eq:balanceTests}, we observe
that
\[
  l(\vec{\Delta}) \models B \implies
  (\Delta_1=\delta_1) \models B_1 \implies
  \delta_1 = - \Delta_S.
\]
When $\delta_1 \neq -\Delta_S$, it implies that
$$
 \Prc{l(\vec{\Delta}) \models B}{\vec{V}_k \wedge (\Delta_1=\delta_1)} = 0,
$$
which makes all the terms in the summation vanish but the term
$\Delta_1=-\Delta_S$. Applying the above introduced shorthand $l_1$,
the transition probability is equal to
\begin{equation}
  \label{eq:l1}
  \Prc{l(\vec{\Delta}) \models B}{\vec{V}_k \wedge l_1}
  \cdot
  \Prc{l_1}{\vec{V}_k}.
\end{equation}

We can then iterate the application of the total probability law on
the first term of \eqref{eq:l1} as follows
\[
  \begin{aligned}
    & \sum_{(\delta_2,\delta_3,\delta_9) \in D^3}
    \!\!\!\!\!\!\!\!\!\!\Prc{l(\vec{\Delta}) \models B}{\vec{V}_k \wedge l_1 \wedge l_5}
     \cdot \Prc{l_5}{\vec{V}_k \wedge l_1 }.
  \end{aligned}
\]
We can again observe that the triple summation is only apparent, since
for the terms that do not satisfy $B_2$, that is,
$(\Delta_1=\delta_1, l_5) \nvDash B_2 $, it holds that
\[
  \Prc{l(\vec{\Delta}) \models B}{\vec{V}_k \wedge l_1 \wedge
    l_5} = 0.
\]
As in the previous case, considering the second balance equation
\eqref{eq:balanceTests}, we have introduced the shorthand $l_2$, which
simplifies the notation for the above term to:
\[
\begin{aligned}
& \Prc{l(\vec{\Delta}) \models B}{\vec{V}_k \wedge l_1 \wedge
    l_5} = \\
& \sum_{\delta_2=0}^{-\Delta_S-\Delta_A-\Delta_{R_a}}
  \!\!\!\!\!\!\!\!\!\!\!\!\!\!\Prc{l(\vec{\Delta}) \models B}{\vec{V}_k \wedge l_1  \wedge l_2 }
   \Prc{l_2}{\vec{V}_k \wedge l_1 }.
\end{aligned}
\]
This reasoning can be iterated for all the nested conditional probabilities until the
expression in the claim is obtained.
\end{proof}

In the following sections we will report the computations of the
different probabilities that appear in~\eqref{eq:transProbFinalExt}.

\subsection{Probability that a susceptible becomes infectious}
\label{subsec:ProbInfected}

A first step towards the computation of the different components
of~\eqref{eq:transProbFinalExt} is to compute the probability that a
susceptible subject becomes infectious.  To this end, we first
introduce the following useful auxiliary events (the negated events
are in parentheses) taking place at time $k$
\begin{equation}\renewcommand{\arraystretch}{0.9}
  \hspace*{-3mm}\begin{dcases*}
    c_k \, (\overline{c}_k) & \hspace*{-2mm}\parbox[t]{5.5cm}{A subject meets one person and (does not) contracts the infection},\\
    m_k \, (\overline{m}_k) & \hspace*{-2mm}\parbox[t]{5.5cm}{A subject meets one person and he/she is (not) infectious },\\
    g_k \, (\overline{g}_k) & \hspace*{-2mm}\parbox[t]{5.7cm}{A person (does not) becomes infected after meeting an infectious person}.\\
  \end{dcases*}
\end{equation}
Hence, the event $g_k$ can be defined as
$g_k = \left\{p \in \mathcal{S}_k \wedge p \in \mathcal{A}_{k+1}
\right\}$,
where $p$ is a generic person. Since each subject has $M$ different
meetings during time $k$, we have
\[
  \Prc{g_k }{\vec{V}_k}_M = 1 - \Prc{\overline{c}_k}{\vec{V}_k}^{M} ,
\]
i.e., the probability of contracting the infection is the
complementary of the probability of remaining healthy after $M$
meetings.  It can be observed that a person remains healthy after
one meeting either if the person met is herself healthy or if she is
infectious but the infection is not transmitted in the meeting, i.e.
$\overline{c}_k = \overline{m}_k \cup (m_k \cap \overline{g}_k)$.
Thereby,
\begin{align*}
  &\Prc{\overline{c}_k}{\vec{V}_k} = \Prc{\overline{m}_k \cup (m_k \cap \overline{g}_k)}{\vec{V}_k}\\
  &=\Prc{\overline{m}_k}{\vec{V}_k} + \Prc{m_k \cap \overline{g}_k}{\vec{V}_k}\\
  &= \Prc{\overline{m}_k}{\vec{V}_k} + \Prc{\overline{g}_k}{m_k\wedge
    \vec{V}_k} \Prc{m_k}{\vec{V}_k}.
\end{align*}
Assuming a uniform distribution in the meetings across the different
classes of subjects and recalling that
$\Prc{\overline{g}_k}{m_k\wedge \vec{V}_k} = 1 -\omega$, we have:
\[
\begin{aligned}
  \Prc{\overline{c}_k}{\vec{V}_k} &= \frac{S_k+R_k+R_{a_k}}{N-D_k-I_k-O_k-Q_k}\\
  & \quad +\frac{A_k(1-\omega)}{N-D_k-I_k-O_k-Q_k}\\
  &= 1 - \omega\frac{A_k}{N-D_k-I_k-O_k-Q_k},
\end{aligned}
\]
where we applied~\eqref{eq:domain_invariantTested}.
Therefore
\begin{equation}
  \label{eq:ProbInfectedTested}
\begin{aligned}
  \!\!\!\!\Prc{g_k}{\vec{V}_k}_M\! =\! 1 \!-\! \left( \!1\! - \!
    \frac{\omega A_k}{N\!-\!D_k\!-\!I_k\!-\!O_k\!-\!Q_k} \right)^{M}\!\!\!\!\!\!.
\end{aligned}
\end{equation}

\subsection{Probability that a number of infected subjects are
  identified and quarantined}

In our setting, $t_k$ tests are administered to a set of subjects in
the set
$\mathcal{N}_k = \mathcal{A}_k \cup \mathcal{S}_k \cup
\mathcal{R}_{a_k}$, i.e. on the population that is not in the known
set
$\mathcal{Q}_k \cup \mathcal{I}_k \cup \mathcal{R}_k \cup
\mathcal{D}_k \cup \mathcal{O}_k$.\\
Let the event $s_{H}$ be defined as: ``$H$ people are tested positive,
given the state $\vec{V}_k$''.  Assuming that each subject can be
tested only once at time $k$ (which implies
$t_k \leq N_k = |\mathcal{N}_k|$), our goal is to find the
probability 
$\Pr{s_{H}}$.  To this end, we first recall~\eqref{eq:BinomialPMF}
and, by defining the event $a_p$ ``$p$ infected subjects are tested'',
we immediately have that the probability that $H$ subjects are found
positive given that $p\in\{0,\dots,A_k\}$ infected subjects are
tested, is given by
\begin{equation}
  \label{eq:ProbTestPos}
  \Prc{H}{a_p} = \mathcal{B}(p,\gamma)_H .
\end{equation}
We are now in a position to prove the following theorem.

\begin{theorem}
  Given $t_k \leq N_k = A_k+S_k+R_{a_k}$ \mr{and $H \leq t_k$}, we have that
  \[
    \Pr{s_{H}} = \binom{N_k}{t_k}^{-1} \sum_{p = 0}^{t_k} \binom{S_k
      + R_{a_k}}{t_k - p} \binom{A_k}{p} \mathcal{B}(p,\gamma)_H .
  \]
\end{theorem}

\begin{proof}
  The probability $\Pr{s_{H}}$ is given by the sum of disjoint
  events (Total Probability Law) that enumerates all the possibilities
  of testing at most $t_k$ infected in the set $\mathcal{A}_k$, i.e.
  $\Pr{s_{H}} = \sum_{p=0}^{t_k}\Pr{a_p} \Prc{H}{a_p}$,
  where $\Prc{H}{a_p}$ is given in~\eqref{eq:ProbTestPos}.  To
  determine $\Pr{a_p}$, we first consider the number $p$ as the
  ``number of infected people tested at time $k$''. Hence, we have
  that the probability of one possible sequence having exactly $p$
  infected people tested is equal to: the number of the different ways
  of arranging $t_k - p$ people from the negative set of people
  $\mathcal{S}_k \cup \mathcal{R}_{a_k}$ times the number of the
  different ways of arranging $p$ people from the positive set of
  people $\mathcal{A}_{k}$ divided by the number of the different ways
  of arranging $t_k$ people from the overall set of people $N_k$, i.e.
  \[
    \frac{\mathbb{D}_{S_k+R_{a_k},t_k-p}\mathbb{D}_{A_k,p}}{\mathbb{D}_{N_k,t_k}}
    .
  \]
 This number has to be multiplied by the number of all the
  possible different sequences, which is given by all the
  permutations of $t_k$ elements taken from the two sets with $t_k-p$
  and $p$ elements, i.e., $\binom{t_k}{p}$.
  By substituting the definitions given previously, we have that
  \[
    \begin{aligned}
    \Pr{a_p} &=
    \frac{\mathbb{D}_{S_k+R_{a_k},t_k-p}\mathbb{D}_{A_k,p}}{\mathbb{D}_{N_k,t_k}}
    \frac{\mathbb{P}_{t_k}}{\mathbb{P}_{t_k - p}\mathbb{P}_p} \\
     & = \binom{N_k}{t_k}^{-1} \binom{S_k + R_{a_k}}{t_k - p}
    \binom{A_k}{p} .
        \end{aligned}
  \]
  Hence the proof follows.
\end{proof}





\subsection{Probability of a state change}

Instrumental to the construction of the transition matrix is to
compute the probability that a given number of subjects simultaneously
change their state, i.e. the different factors in
equation~\eqref{eq:transProbFinalExt} of
Theorem~\ref{teo:trasProb}. Considering Figure~\ref{fig:stateTest},
such probabilities are discussed next.

\textbf{Transition from State $S$.}  The probability of the event
``exactly $\Delta_1$ susceptible subjects become infected at step $k$
given the state $\vec{V}_k$ and a choice for $M$'' is approximated by
$\mathcal{B}(S_k, \Prc{g_k}{\vec{V}_k}_M)_{\Delta_1}$.
Hence, the probability of the assignment $l_1$ is approximated by
\begin{equation}
  \label{eq:FirstCombProb}
  \Prc{l_1}{\vec{V}_k} =  \mathcal{B}(S_k,
  \Prc{g_k}{\vec{V}_k}_M)_{-\Delta_{S}}. 
\end{equation}

\textbf{Transition from State $A$.}  Consider the event
$r^{(E)}_{\delta_2, \delta_3, \delta_9}$ defined as ``exactly
$\delta_3$ asymptomatic infected subjects recover at step $k$ without
being quarantined, $\delta_2$ subjects become symptomatic and
$\delta_9$ out of $t_k$ are tested positive and become quarantined
given the state $\vec{V}_k$''.  Let
$\rho^{(E)}(\delta_2,\delta_3, \delta_9, \vec{V}_k)_{t_k}$ be the
probability of this event given $\vec{V}_k$ and $t_k$. Instrumental to
the computation of this probability, the following theorem proves
useful.

\begin{theorem}
  \label{th:RhoTheorem}
  Let $\rho(\delta_2,\delta_3, \vec{V}_k)$ be the probability of the
  event ``exactly $\delta_3$ asymptomatic infected subjects recover at
  step $k$ and $\delta_2$ subjects become symptomatic''.  Hence,
  $\rho(\delta_2,\delta_3, \vec{V}_k) = 0$ if
  $\delta_2+\delta_3 > A_k$, otherwise
  \[
    \rho(\delta_2, \delta_3, \vec{V}_k) =
    \mathbb{M}_{A_k,\delta_2,\delta_3} \beta^{\delta_3}\,
    \delta^{\delta_2}\, C_{\beta, \delta}^{A_k-\delta_2-\delta_3} .
  \]
\end{theorem}
\begin{proof}
  Since $\beta$ is the probability for an infectious asymptomatic
  subject to recover, $\delta$ is the probability for an asymptomatic
  subject to develop symptoms and $C_{\beta, \delta} \geq 0$, as
  defined in Section~\ref{sec:problem}, we have immediately that
  $\rho(\delta_2,\delta_3, \vec{V}_k) = 0$ if
  $\delta_2+\delta_3 > A_k$. Moreover, the probability
  $\rho(\delta_2, \delta_3, \vec{V}_k)$ is given by all the possible
  combinations of $\delta_2$ and $\delta_3$ elements out of the $A_k$
  asymptotic subjects weighted by the respective probabilities, hence
  \[
    \rho(\delta_2, \delta_3, \vec{V}_k) = \frac{A_k!\,
      \beta^{\delta_3}\, \delta^{\delta_2}\, C_{\beta,
        \delta}^{A_k-\delta_2-\delta_3}}{\delta_2! \delta_3!
      (A_k-\delta_2-\delta_3)!} ,
  \]
  which concludes the proof.
\end{proof}

We can now show the probability of the event
$r^{(E)}_{\delta_2, \delta_3, \delta_9}$.

\begin{theorem}
  The probability
  $\rho^{(E)}(\delta_2,\delta_3, \delta_9, \vec{V}_k)_{t_k}$ is zero
  in the trivial \dan{conditions
    $\delta_2+\delta_3+\delta_9 > A_k \vee \delta_9 > t_k \vee
    \delta_2 < H - \delta_9$}, and otherwise is given by \dan{%
    \begin{equation}
    \label{eq:rhoETested}
    \begin{array}{@{}l@{}}
    \rho^{(E)}(\delta_2,\delta_3, \delta_9, \vec{V}_k)_{t_k} = \\
      \quad \displaystyle \sum_{H = \delta_9}^{t_k} \Pr{s_H} \sum_{F = 0}^{\delta_9} \rho(\delta_2, \delta_3\!+\!F, \vec{V}_k) \binom{A_k}{H}^{-1} K,
    \end{array}
  \end{equation}
  where
  \begin{equation}
    K = \binom{\delta_3+F}{F} \binom{A_k - (\delta_2 +
      \delta_3 + F)}{\delta_9 - F} \binom{\delta_2}{H - \delta_9} .
  \end{equation}
}
\end{theorem}

\begin{proof}
  Recalling the definition of the event $s_H$ (i.e., $H$ subjects are
  tested positive given the state $\vec{V}_k$), thus, conditioning over
  $s_H$, we have:
  \[
    \begin{aligned}
    \rho^{(E)}(\delta_2,\delta_3, \delta_9, \vec{V}_k)_{t_k} &=
    \Pr{r^{(E)}_{\delta_2, \delta_3, \delta_9}} \\
    &= \sum_{H=0}^{t_k}
    \Pr{s_H} \Prc{r^{(E)}_{\delta_2, \delta_3, \delta_9}}{s_H},
    \end{aligned}
  \]
  where the first equality follows by definition.
  Introduce the event $r_{\delta_2, \overline{\delta}_3}$ described as
  ``exactly $\overline{\delta}_3$ asymptomatic infected subjects
  recover at step $k$ and $\delta_2$ subjects become symptomatic given
  the state $\vec{V}_k$'', which, according to Theorem~\ref{th:RhoTheorem}, has probability
  $\rho(\delta_2, \overline{\delta}_3, \vec{V}_k)$.  It is important to observe that the
  $\delta_2$ subjects will entirely move to state $I$, while among
  the $\overline{\delta}_3$ subjects, $\delta_3$ that are not tested
  positive move to the state $R_a$, whilst
  $F = \overline{\delta}_3 - \delta_3$ are tested positive and move to
  the state $Q$. The number of the subjects $F$ that recover and are
  tested positive ranges from $0$ to
  $\min(\overline{\delta}_3,\delta_9)$, with $\delta_9$ the positive tested out of $t_k$.  Thereby,
  \begin{align*}
     & \Prc{r^{(E)}_{\delta_2, \delta_3, \delta_9}}{s_H} \\
     & \quad = \sum_{F=0}^{\delta_9} \Prc{r^{(E)}_{\delta_2, \delta_3,
        \delta_9}}{s_H \!\wedge\! r_{\delta_2, \delta_3\!+\!F} }
    \Prc{r_{\delta_2, \delta_3+F}}{s_H} .
    \end{align*}
  The event $r_{\delta_2, \delta_3+F}$ is independent from the event
  $s_H$, since testing and healing are independent, hence
  \[
    \begin{aligned}
      \Prc{r_{\delta_2, \delta_3+F}}{s_H} &= \Pr{r_{\delta_2, \delta_3+F}} \\
      & = \rho(\delta_2, \delta_3+F, \vec{V}_k) =
      \rho(\delta_2, \overline{\delta}_3, \vec{V}_k).
    \end{aligned}
  \]
  We take into account now the conditional probability
  $\Prc{r^{(E)}_{\delta_2, \delta_3, \delta_9}}{s_H \wedge
    r_{\delta_2, \delta_3+F} }$ that considers that $H$ subjects are
  tested positive, we notice that: a) $F$ of them are subjects that
  would recover (if they did not test positive) and are extracted from
  a group of $\overline{\delta}_3 = \delta_3+F$ subjects,
  i.e. $\frac{\mathbb{D}_{\delta_3+F,F}}{\mathbb{P}_{F}}$; b) since a
  number of $A_k - (\delta_2 + \delta_3 + \delta_9)$ remains in the
  asymptomatic state \dan{$\mathcal{A}$}, $\delta_9 - F$ are subjects
  that would remain asymptomatic and are extracted from a group of
  $A_k - (\delta_2 + \overline{\delta}_3)$ subjects, i.e.
  $\frac{\mathbb{D}_{A_k - (\delta_2 + \delta_3 + F),\delta_9 -
      F}}{\mathbb{P}_{\delta_9 - F}}$; c) $H - \delta_9$ are subjects
  that develop symptoms and are extracted from a group of $\delta_2$
  subjects, i.e.
  $\frac{\mathbb{D}_{\delta_2,H-\delta_9}}{\mathbb{P}_{H -
      \delta_9}}$.  \dan{Therefore, by dividing by all the possible ways
    to arrange $H$ elements tested positive in $\mathcal{A}$, we have
\begin{align*}
    & \Prc{r^{(E)}_{\delta_2, \delta_3, \delta_9}}{s_H \wedge
      r_{\delta_2, \delta_3+F} } \\
    & \quad  =  \frac{\mathbb{D}_{\delta_3+F,F}}{\mathbb{P}_{F}}
    \frac{\mathbb{D}_{A_k - (\delta_2 + \delta_3 + F),\delta_9 -
        F}}{\mathbb{P}_{\delta_9 - F}}
    \frac{\mathbb{D}_{\delta_2,H-\delta_9}}{\mathbb{P}_{H -
        \delta_9}} \frac{\mathbb{P}_{H}}{\mathbb{D}_{A_k,H}},
\end{align*}
}
  and the proof follows.
\end{proof}

The probability $\Prc{l_2}{\vec{V}_k \wedge l_1 }$ is therefore given
by:
\begin{equation}\label{eq:l2e}
  \Prc{l_2}{\vec{V}_k \wedge l_1 } = \rho^{(E)}(\delta_2,\delta_3, \delta_9, \vec{V}_k) ,
\end{equation}
with \mr{$\rho^{(E)}(\cdot)$} \ given by equation~\eqref{eq:rhoETested}.


\textbf{Transition from State $I$.} Consider the event ``exactly
$\delta_4$ symptomatic subjects recover, $\delta_5$ become
hospitalised and $\delta_6$ die''.  Let
$\phi(\delta_4, \delta_5, \delta_6, \vec{V}_k)$ be its probability
given $\vec{V}_k$.  The probability
$\phi(\delta_4, \delta_5, \delta_6, \vec{V}_k)$ is zero if
$\delta_4+\delta_5+\delta_6 > I_k$ or if $\delta_5 > C-O_k$ (the
latter condition meaning that we are hospitalising more patients than
allowed by the capacity of the hospital).  We have two cases, when
$\delta_5 < C-O_k = \overline{C}_k $, that is, we can hospitalise the
patients, and when the maximum available capacity is saturated:
\begin{equation}
\begin{aligned}
  &\!\!\!\phi(\delta_4,\delta_5, \delta_6, \vec{V}_k) =\!\\
  & = \!\!\begin{cases}
0 & \!\!\!\text{if }\delta_4+\delta_5+\delta_6 > I_k \vee \delta_5 > \overline{C}_k \\
M(\delta_4, \delta_5, \delta_6) & \!\!\!\text{if } \delta_5 < \overline{C}_k \wedge \delta_4+\delta_5+\delta_6 \leq I_k \\
M'(\delta_4, \delta_6)  & \!\!\!\text{if } \delta_5 = \overline{C}_k \wedge \delta_4+\delta_5+\delta_6 \leq I_k
\end{cases}\!\!
\end{aligned}
\label{eq:phiComp}
\end{equation}
where, similar to Theorem~\ref{th:RhoTheorem}, we have
\[
  M(\delta_4, \delta_5, \delta_6) =
  \mathbb{M}_{I_k,\delta_4,\delta_5,\delta_6} \mu^{\delta_4}\,
  \psi^{\delta_5}\, \alpha^{\delta_6}\, C_{\mu, \psi, \alpha}^{I_k -
    \delta_4 - \delta_5 - \delta_6} .
\]
The idea is that whenever the result for the
hospitalised would exceed the maximum residual capacity
$\overline{C}_k$, the extra patients are turned down and remain in
state $I$.  Thereby,
\[
M'(\delta_4, \delta_6) =
\sum_{h = 0}^{I_k - \delta_4 - \delta_6 -\overline{C}_k} M(\delta_4, \overline{C}_k+h, \delta_6),
\]
where we are accumulating into $M'(\delta_4, \delta_6)$ the
probability of having more patients evaluated as hospitalisable than
it is possible to receive.
We can now compute the probability
\begin{equation}
  \label{eq:l3e}
 \Prc{l_3}{\vec{V}_k \wedge l_1 \wedge l_2} = \phi(\delta_4, \delta_5, \delta_6, \vec{V}_k) .
\end{equation}

\textbf{Transition from State $O$:} Consider the event ``exactly
$\delta_8$ hospitalised subjects recover at step $k$ and $\delta_7$
hospitalised subjects die'' and let
$\zeta(\delta_7,\delta_8, \vec{V}_k)$ be its probability,
given $\vec{V}_k$, which is obviously zero if
$\delta_7+\delta_8>O_k$. The remaining cases show again a
multinomial coefficient, i.e.,
\begin{equation}\label{eq:zetaF}
  \zeta(\delta_7,\delta_8, \vec{V}_k)
  = \mathbb{M}_{O_k,\delta_7,\delta_8}
  \sigma^{\delta_7}\, \xi^{\delta_8}\,
  C_{\sigma,\xi}^{O_k-\delta_7-\delta_8}.
\end{equation}

\textbf{Transition from State $Q$.}  The transitions from state $Q$ is
similar to the previous case from state $O$, hence it is given by
\begin{equation}\label{eq:chiF}
  \chi(\delta_{10},\delta_{11}, \vec{V}_k)
  =
  \mathbb{M}_{Q_k,\delta_{10},\delta_{11}} \iota^{\delta_{10}}\,
  \upsilon^{\delta_{11}}\,
  C_{\iota,\,\upsilon}^{Q_k\!-\!\delta_{10}\!-\!\delta_{11}}.
\end{equation}
We can now compute the term
$\Prc{l_4}{\vec{V}_k \wedge l_1 \wedge l_2 \wedge l_3}$ that connects
together the transitions from sates $O$ and $Q$:
\begin{equation}
  \label{eq:l4e}
  \scriptsize
  \begin{array}{@{}l@{}}
    \!\!\Prc{l_4}{\vec{V}_k \wedge l_1 \wedge l_2 \wedge l_3} =\\
    \;\;= \zeta(\Delta_D\! -\! \delta_6, \delta_5\! +\!\delta_6\! -\! \Delta_D \!-\! \Delta_O, \vec{V}_k) \\
    \;\;\cdot \chi(\Delta_I \!-\! \delta_2\!+\!\delta_4\!+\!\delta_5\!+\!\delta_6, \Delta_R \!+\! \Delta_D \!+\! \Delta_O \!-\! \delta_4 \!-\!\delta_5 \!-\! \delta_6, \vec{V}_k).\!
  \end{array}
\end{equation}

The computation of the transition probabilities defined in
Theorem~\ref{teo:trasProb} is thus obtained by
plugging~\eqref{eq:FirstCombProb},~\eqref{eq:l2e},~\eqref{eq:l3e}
and~\eqref{eq:l4e} into ~\eqref{eq:transProbFinalExt}.

\dan{\subsection{Complexity analysis}

  For the analysis of the complexity, we will consider the overall
  number of states that can be generated by the model, assuming an
  arbitrary number of model configurations $n$, i.e., susceptible,
  asymptomatic, etc. Let $N$ be the subjects in the population, we may
  assume that the overall number of states $m$ that can be generated
  is given by $m = (N + 1)^n$. However, this is a large upper bound to
  the states that can be actually reached: indeed, given $\alpha_i$,
  $i = 1,\dots,n$, the number of subjects in the $i$-th configuration,
  we have the constraint $\sum_{i=1}^n \alpha_i = N$. To account for
  this simplifying constraint, we first define $m = f(n,N)$, i.e., the
  overall number of possible states $m$ given the $n$ configurations
  and $N$ subjects. The number $m$ can be computed considering that
  the $n$-th configuration accounts for an arbitrary number
  $0 \leq i \leq N$ of subjects, while the remaining population of
  $N-i$ are allocated in the other $n-1$ configurations. Therefore%
\begin{align*}
  m = f(n, N) = \sum_{i=0}^N f(n-1, N-i)
\end{align*}
Hence, by simply considering that $f(1,N) = 1$, we have%
\begin{align*}
f(2,N) & = N+1, \\
f(3,N) & = \frac{(N+2)(N+1)}{2} = \binom{N+2}{2} = \binom{N+2}{N},
\end{align*}
we end up with
\begin{align*}
  m = f(n,N) = \binom{N+n-1}{n-1} = \binom{N+n-1}{N},
\end{align*}
which is the Bose-Einstein statistics. Since the complexity is
proportional to the number of reachable states, we have immediately
the complexity figure for the presented algorithm.}


\section{Numeric Examples}
\label{sec:numeric}
To show the potential applications of the formalised model, we have
developed in C++ \mr{a proof of concept} open-source tool.\footnote{The
  tool and all the information to reproduce these experiments are
  available from \url{https://bit.ly/33jyiZU}.
}
\mr{The tool operates both in the case $t_k$ is set to zero (i.e., no
  people ever get quarantined) or not. For the rest of the section we
  consider the case where $t_k = 0$.}
  In this simplified situation we could generate and
analyse the Markov model for populations of size comparable to a
typical circle of people who meet on a regular basis. What is more,
the tool can be interfaced with the \texttt{PRISM} stochastic model
checker~\cite{kwiatkowska2011prism}.
%
The analysis of a system goes through the following steps.
\begin{enumerate*}[label=\arabic*.]
\item reachability analysis: starting from an initial set of states
  and from a complete specification of the system's parameters (e.g.,
  the different probabilities, the population size), compute all the
  states that are reachable and construct a transition table that
  associates each \mr{state-action pair \texttt{(cS, a)} with the next
    state \texttt{nS} and probability \texttt{p} to reach that state,
    where the \texttt{a} is associated with the number $M$ of people
    that are allowed to be met for that state \texttt{cS}.}
\item application of a \texttt{Policy}: apply a control law
  associating each state with one of the actions that are possible for
  that state generating a DTMC.
\end{enumerate*}
After the completion of the second step, it is possible to either
solve the DTMC (i.e., given a set of initial states and a known initial
probability, compute the evolutions for different steps and/or at the
steady-state), or apply the PRISM model checker to verify some
Probabilistic Linear Temporal Logic (PLTL) properties~\cite{HJ94}.\\
Hereafter, we will use numeric examples to expose some interesting
facts related to the properties of the formalised model. We used the
following choice of probabilities
($\alpha$,$\beta$,$\delta$,$\mu$,$\omega$,$\psi$,$\sigma$,$\xi$)=(0.25,0.45,0.25,0.4,0.5,0.35,0.1,0.65).
%
\mr{We executed the experiments on a PCs equipped with 256Gb of RAM,
  Intel\textsuperscript{\textregistered}
  Xeon\textsuperscript{\textregistered} 2.20GHz with 12 cores running
  Linux.}

\begin{figure}[t]
\centerline{\includegraphics[width=0.95\columnwidth]{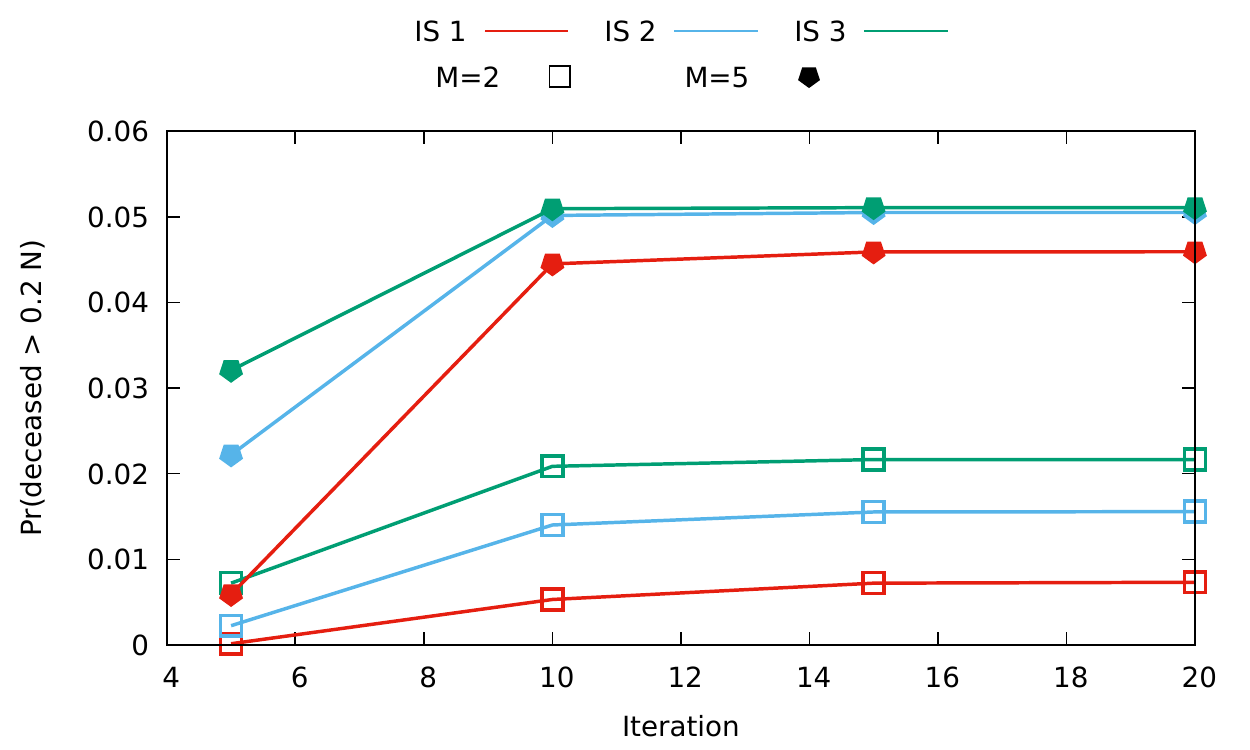}}
\caption{\label{fig:ProbVsM}Probability that more than 20\% of the population will die as a function of the iterations for two different values of $M$.}
\end{figure}
\begin{figure}[t]
  \centering
  \includegraphics[width=0.95\columnwidth]{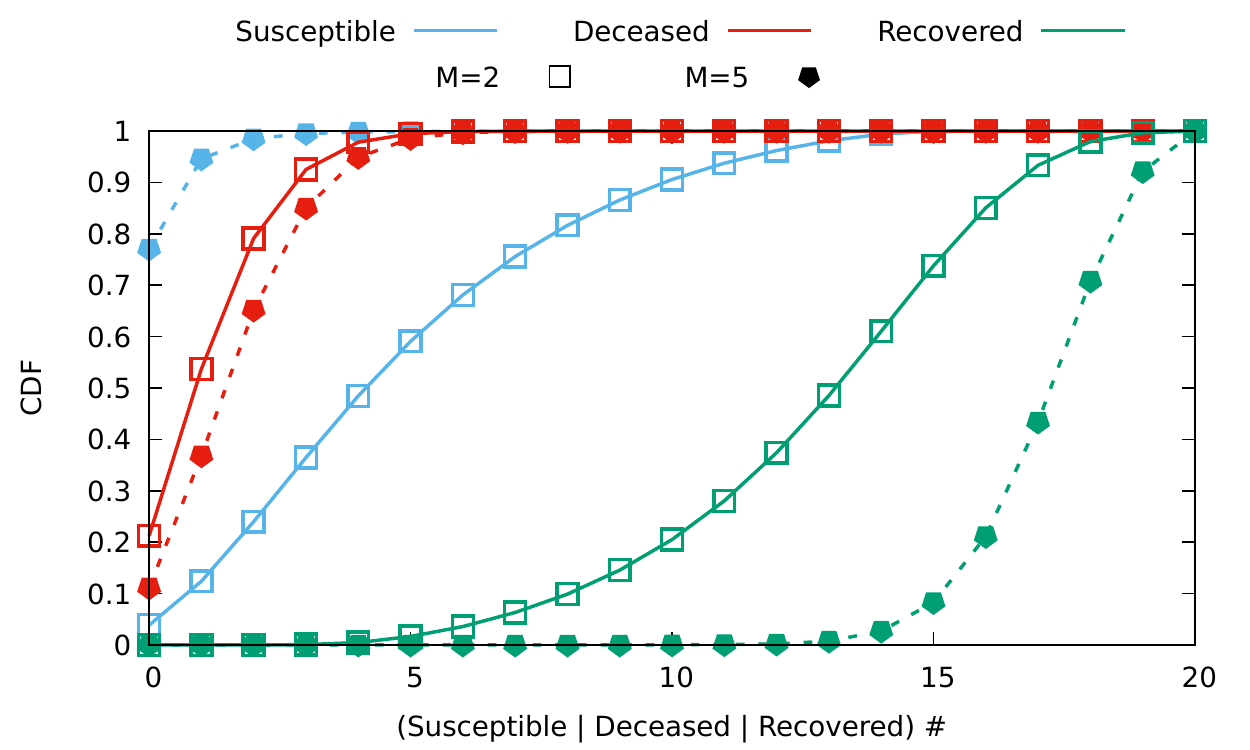}
  \caption{\label{fig:CdFVsM}Steady-state CDFs of the deceased, the recovered and of the susceptible subjects for initial state 3 and variable $M$.}
\end{figure}

\noindent
\textbf{Equilibrium and Convergence.}
When we apply a policy, the system reduces to a DTMC. The presence of
multiple absorbing states prevents the existence of a single
steady-state equilibrium (trivially, any initial state such that all
elements are zero but the absorbing states is an equilibrium
point).
The initial states worth to consider are those where most of the
subjects are susceptible and a few of them are asymptomatic and they
start spreading the infection.
In this case, our question is to compute the probability that no more
than a given percentage of the population (e.g., $20\%$) will
eventually die owing to the virus.
We considered a population of $N=20$ subjects with the following three
initial states:
%
\begin{enumerate*}[label=\bfseries IS \arabic*]
\item = (19,1,0,0,0,0),
\item = (17,3,0,0,0,0),
\item = (15,5,0,0,0,0).
\end{enumerate*}
%
%
Each initial state is represented with a tuple
$(S_0,A_0,I_0,R_0,O_0,D_0)$, where each element corresponds to the
population in the respective state at step $0$.
In order to emphasise the impact of the choice of $M$ on the
equilibrium, we fixed parameter $C$ (i.e., the capacity of the
intensive care facility) to a relatively high value: $C = 5$.  The
result of the benchmark probability
$\lim_{k\rightarrow \infty} \Pr{D_k \geq 0.2 N}$ is reported in
Figure~\ref{fig:ProbVsM}.
The figure reveals that the system converges to its equilibrium in a
relatively small number of iterations (around $10$). Another important
fact is that the choice of a different initial state produces a
different steady-state probability. Interestingly, these probabilities
are closer for a permissive policy ($M=5$), while they spread apart
for a more restrictive policy ($M=2$). The steady state Cumulative
Distribution Functions (CDF) cast some light on this phenomenon.  In
Figure~\ref{fig:CdFVsM}, we show the steady marginal distribution for
the IS 3 and for two different values of $M$.
%
%
The figure shows that aggressive restrictions gradually decrease
the number of people in the asymptotic state without ``fresh''
replenishment from the susceptible state. As a consequence, at some
point, with high probability, no more subjects will be infected, while
the infected people will move to one of the absorbing state (deceased
or recovered).  On the contrary, a permissive policy causes all the
subjects to be quickly infected.  The consequences on the deceased are
not substantial in this case because for this experiment, it is
assumed to be a ``generous'' availability of intensive care beds,
but the difference is visible on the number of subjects that get the
disease and recover.  When a high number of subjects remains confined
to the susceptible state, the steady state becomes highly dependent on
how many subjects are initially infected (which is the reason why for
$M=2$ the steady state probability is highly dependent on the initial
state).

\begin{figure}[t]
\centerline{\includegraphics[width=0.95\columnwidth]{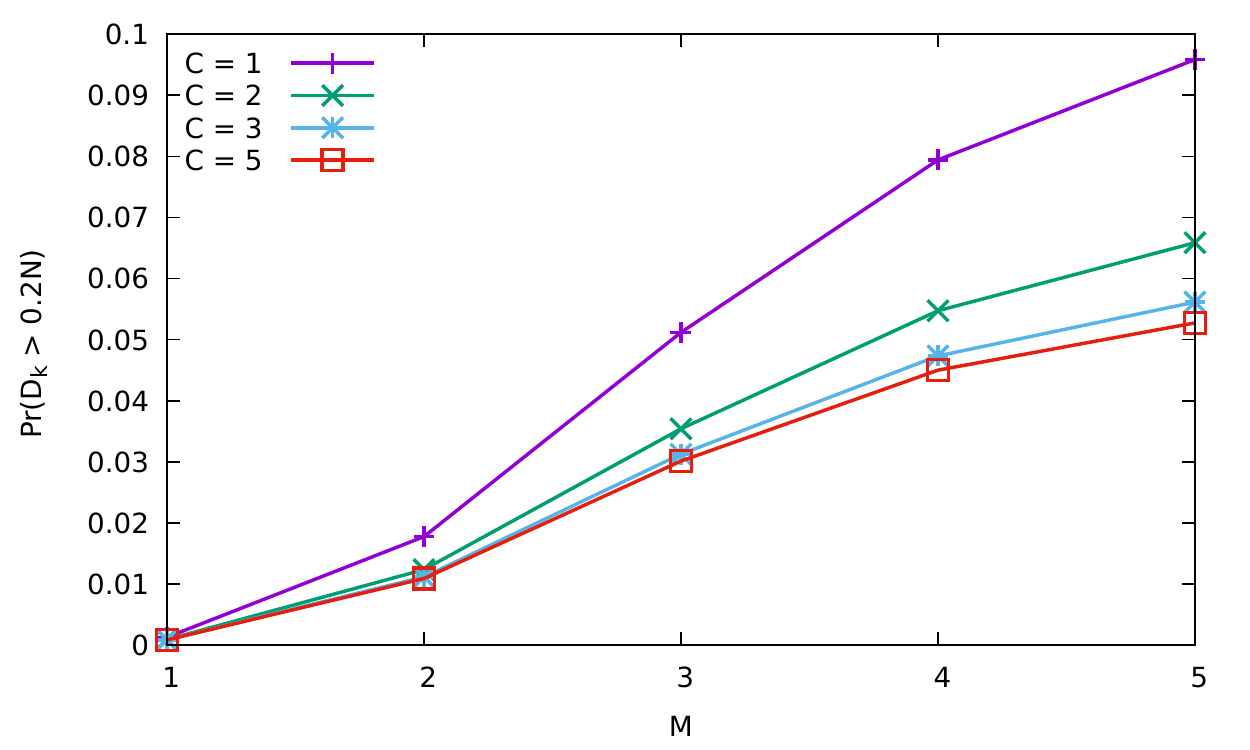}}
\caption{\label{fig:ProbVsMC}Probability that more than 20\% of the population will die as a function of $M$ for different capacities $C$.}
\end{figure}

\textbf{Impact of Social Distancing.}
\label{sec:socialDist}
%
In order to design adaptive policies, it is of paramount importance to
look at the impact of the input variable $M$, i.e., the number of
people that each one is allowed to meet at each step.  For this
experiment we assumed that at time $0$ the system could be either in
(15,3,0,0,0) or in (17,1,0,0,0) %
%
%
with equal probability $0.5$, and the capacity of the intensive care
$C$ was chosen \mr{in the set $\{1, 2, 3, 5\}$}. The plot of the probability of having
more than $20\%$ casualties is shown in Figure~\ref{fig:ProbVsMC}.
The dependence is monotonic in $M$, but the amount of change clearly
depends on the hospital capacity $C$. For $C=1$ and $M=5$, this probability is beyond $9\%$. Increasing the number of beds in the hospital
has obvious benefits, but for the system at hand the performance gain
flattens for $C \geq 3$, meaning that additional growths in the beds
are not worth the effort.

\begin{figure*}
\centering
\begin{tabular} {ll}
\includegraphics[width=0.85\columnwidth]{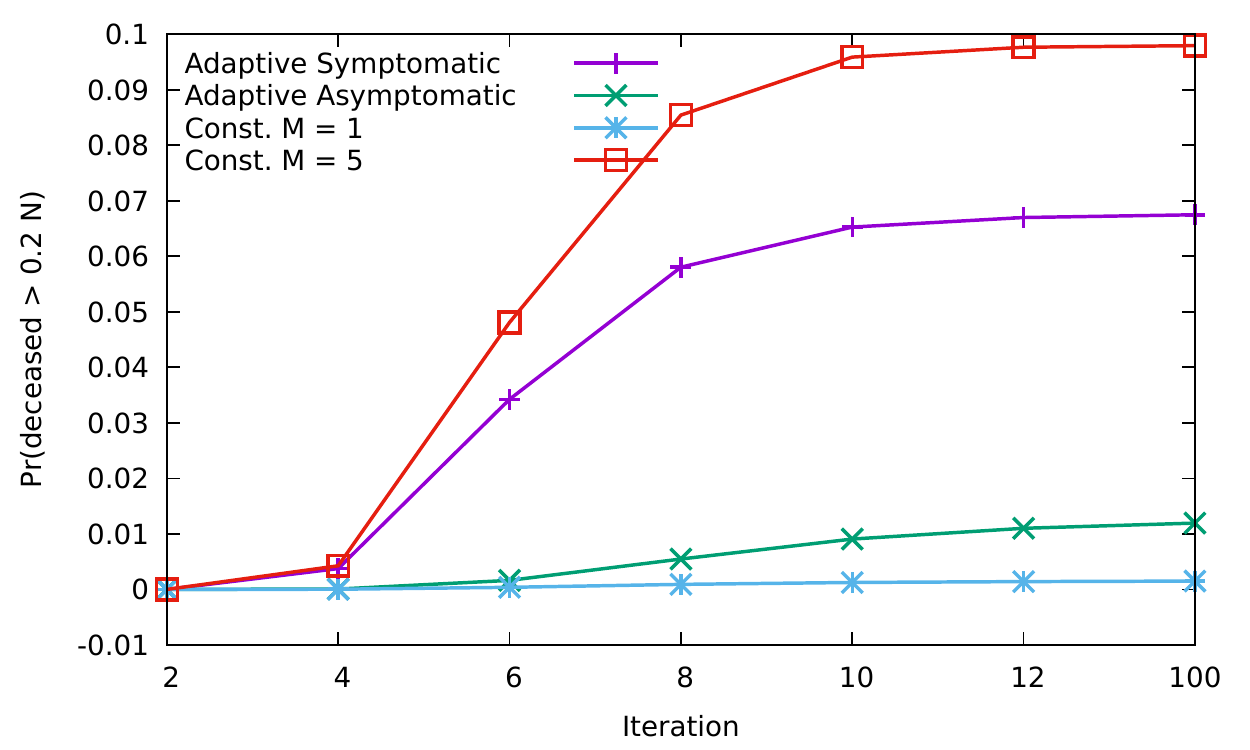} &
\includegraphics[width=0.85\columnwidth]{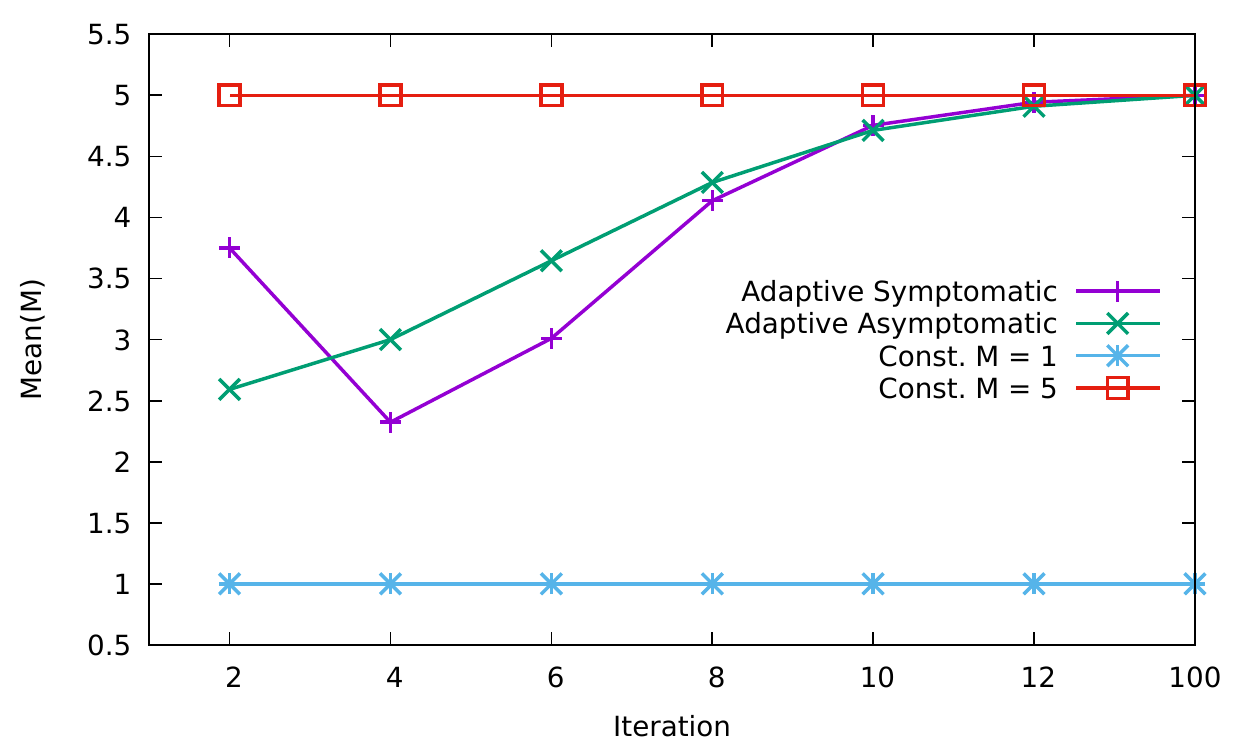}\\
\end{tabular}
\caption{\label{fig:adaptivePlot}Probability that more the $20\%$ of the population will die, average value of the action $M$ used in the different iterations.}
\end{figure*}

\textbf{Adaptive Policies.}
An adaptive policy is a feedback control law that decides the social
distance $M$ based on the current value $\vec{V}_k$.
Adaptive policies for a system like the one described in this paper
can be synthesised using the theory of Partially Observable Markov
Decision Processes (POMDPs)~\cite{feinberg2012handbook,
  kaelbling1998planning}.
The efficient synthesis of a POMDP tailored to the model is far from
obvious and is reserved for future work. In this section, we opt for a
simpler choice and present two heuristics, both based on the following
equation:
\begin{equation}
\mathcal{M}(\vec{V}_k)\! =\! \begin{cases}
  M_{\uparrow}   \quad\quad\quad \text{if } f(\vec{V}_k) \leq T_{\downarrow}\\
  M_{\downarrow} \quad\quad\quad \text{if } f(\vec{V}_k) \geq T_{\uparrow}\\
  M_{\uparrow}\! +\! \left(M_{\downarrow}\!-\!M_{\uparrow}\right) \frac{f(\vec{V}_k)-T_{\downarrow}}{T_{\uparrow}-T_{\downarrow}} \text{ otherwise}.\!\!
  \end{cases}\!\!
\end{equation}
The two defined heuristics differ for the choice of the function
$f(\vec{V}_k)$.
The first one set $f(\vec{V}_k)= I_k + O_k/(N-D_k)$ and the rational
is as follows. When the percentage of the number of symptomatic and
hospitalised patients over the living population is below the
threshold $T_{\downarrow}$, we are allowed to use the maximum value
$M_{\uparrow}$ for $M$ (meaning the least restrictions for social
life). If this number is above a threshold $T_{\uparrow}$, we are
forced to adopt the maximum restriction (the minimum value
$M_{\downarrow}$ for $M$). If the number is in between the two
thresholds we perform a linear interpolation between the minimum and
the maximum values of $M$.
The second heuristic sets $f(\vec{V}_k) = A_k/(N-D_k)$, i.e., we make
the choice as for the first option but this time considering the ratio
between asymptomatic infected subjects and the living population.
Clearly, it is not possible to implement this policy (although $A_k$
could be estimated by statistic means using tests), but still we
believe it is useful to have it as a performance baseline.
We performed a set of tests, considering the same scenario as in
Section~\ref{sec:socialDist} assuming the worst-possible case scenario
for hospital beds: $C=1$.
The parameters for the adaptive policies were chosen as
$T_{\downarrow} = 0.05$, $T_{\uparrow} = 0.15$, $M_{\downarrow}=1$
and $M_{\uparrow}=5$. For the sake of completeness, we also compared
the performance of the adaptive policies setting $M$ to a
constant value (the minimum and the maximum).
The results are shown in Figure~\ref{fig:adaptivePlot}: on the left we
report the benchmark probability $\Pr{D_k \geq 0.2 N}$ for the
different iterations; on the right we show the value of the average
$M$ (i.e., $\sum_{\vec{V}_k} \mathcal{M}(\vec{V}_k) \Pr{\vec{V}_k}$).
%
%
The \textit{Adaptive Asymptomatic (AA)} policy using asymptomatic
subjects shows an excellent performance: the probability of the
deceased subjects is very close to the one for $M=1$, but the average
value of $M$ is higher (it starts from $2.5$ and tends to $5$ after a
few iterations).
The \textit{Adaptive Symptomatic (AS)} policy using symptomatic
subjects has a worse performance in terms of probability, and in the
average applies more social restrictions (smaller $M$).
This happens since the \textit{AA} policy anticipates the restrictions
at the first two steps, while the \textit{AS} policy defers the
decision after it can observe a growth of symptomatic subjects, too
late to prevent many deaths.\\
\mr{We also run $1000$ Monte Carlo simulations with a depth of 100 on
  a population of $100$ individuals proportionally scaling $M$, the
  hospital capacity, and the individuals in the different states in
  the initial conditions.
  The $\Pr{D\textsubscript{k} > 0.2 N}$ is $0.0143$ for AS, $0.0091$
  for AA, $0.0084$ for the constant policy with $M=5$, and $0.0143$
  for the constant policy with $M=25$.
%
%
%
  Although the probabilities are lower than for the case $N=25$, the
  results shows a similar trend.}


\subsection{Analysis using Probabilistic Model Checking}
%

\noindent
\textbf{Background on Probabilistic Model Checking (PMC).}
PMC~\cite{FKNP11} is an effective formal verification
method~\cite{DBLP:books/daglib/0020348} used for analysing stochastic
transitions systems such as MDP, DTMC or CTMC. The analysis is
performed by verifying whether the stochastic model of the system (in
our case a policy) meets the requirements encoded in PLTL~\cite{HJ94},
which allows to express very rich properties. For instance,
$\PProb_{< 0.25} [ \LEventually O_k = C ]$ means the probability of
reaching a state where the intensive care facility is
fully utilised is less than 0.25.
%
%
%
%
$\PProb_{=?} [ \LGlobally \LEventually O_k < C ]$ is an example of
a quantitative property, PRISM computes the probability of
visiting infinitely often states where the intensive
care facility is not fully utilised.
%
%
Intuitively, PMC amounts to automatically check whether all the
computations starting from a given set of initial states of an MDP,
DTMC or CTMC satisfy a given PLTL formula.
As result of the check, PMC produces two kinds of outputs: a) qualitative
answers to indicate whether the property holds or is violated (in this
case generating a witness of the violation), b) quantitative answers to
indicate the probability of satisfaction of the given PLTL formula.
Moreover, PMC allows to compute the vector of steady-states with the
corresponding probabilities, compute the transient probabilities after
a given time, as well as compute the probability a certain condition
holds in the steady-states.\footnote{For the definitions of PLTL
  syntax and semantics see \cite{FKNP11,HJ94}, and visit
  \url{http://www.prismmodelchecker.org/}.}

\textbf{Applications of PMC to verify DTMC model.}
We performed a preliminary analysis, using the
\texttt{PRISM}~\cite{kwiatkowska2011prism} model checker, to verify some PLTL
properties on the DTMC model resulting from the application of the
\textit{AA} policy to our model. In this analysis, we considered a
population of $N=10$ persons, and all possible states (i.e. all possible
consistent configurations for $V$)
as initial.
\texttt{PRISM} proved that
\begin{enumerate*}[label=\roman*)]
\item $\PProb_{=?} [ \LGlobally \LEventually O_k < C ]$ holds (the
  computed probability is 1) in about 5.0 seconds;
\item $\PProb_{=?} [ \LGlobally \LEventually O_k = C ]$ is false (the
  computed probability is 0) in about 5.3 seconds;
\item $\PProb_{=?} [ \LEventually O_k = C ]$ holds for about 42\% of
  the initial states, does not hold for 4\% of the initial states, and
  has a probability $p \in (0,1)$ in the remaining 54\% of the initial
  states.
\end{enumerate*}
The last two properties aim at computing the probability to reach a
state where the number of hospitalised is different/equal (resp.) to
the number of available beds in the hospital.
We also asked \texttt{PRISM} to compute the vector of steady-states with the
corresponding probabilities (it took about 133.0 seconds) as well as
the transient probabilities after 1000 time units (about 7.75
seconds).

\mf{ \noindent\textbf{Scaling to large population size.}  These
  experiments have been carried out with a small population.  However,
  we remark that the population can be scaled using quantisation, as
  in~\cite{nasir2020epidemics}. The price to pay in this case is an
  accuracy reduction due to the quantisation error. Indeed,
  approximating a population of $\hat{N}$ with $N$ representatives,
  accounts for $\hat{N}/N$ subjects per representative. Since each
  representative can be either one or zero, we have a maximum
  approximation error of $\hat{N}/(2N)$ subjects.
%
  As an example, modelling with $N=20$ representatives (as discussed
  above) a population of one million (a medium sized city), we can
  estimate a maximum error of $2.5 \cdot 10^{4}$ subjects, which
  corresponds to an error of less than $2.5\%$ of the population,
  i.e., a relative error of $\hat{N}/(2N) \cdot 1/\hat{N}=1/(2N)$.}

\section{Conclusions and future work}
\label{sec:Conclusion}
A new stochastic model for epidemiology has been proposed. Starting
from the classic susceptible, infected, recovered ($SIR$) model, we
have added additional states, i.e. asymptomatic $A$, hospitalised $O$,
dead $D$, recover from asymptomatic $Ra$ and quarantined $Q$, in order
to model the peculiarities of a \covid{}-like infection.  \lui{The key
advantages of the model are: 1. the model is control oriented and is
analytical, therefore it exposes all parameters and decision variables
that make for design and evaluation of control laws; 2. The model
captures in a detailed way the stochastic nature of the virus
transmission and the impact of resource limitations (hospital beds,
number of test swabs).  The price to pay for this level of detail is
the scalability of the analysis.}  For illustrative purposes, we have
shown numeric examples related to the evolution of a small population
and to the application of PMC to assess the validity of PLTL
properties.\\ This work poses the basis for several future works in
different directions.
%
At model level, the command variables $M$ and $t$ (assumed
deterministic in this paper) could be generalised to stochastic
variables with ``controllable'' mean.
Second, the people in need of intensive care who are denied access to
the hospital because of a saturation of the available sets currently
remain in the $\mathcal{I}_k$ set (see~\eqref{eq:phiComp}).
A different possibility is to introduce a new state to express people
in urgent need of care with high probability of death. This choice
will have to be validated by considering the closest match with the
existing data.  Third, our model considers a restricted circle of
people with no interactions with the remaining population.
\lui{We will explore different possibilities to guarantee scalability
in the analysis. A first possibility is to  model a number of groups that evolve ``almost''
independently, except for a few possible interactions that could take
place in mild lockdown situations (e.g., schools, sport activities,
outdoor recreational activities, museums, cinemas, etc.).  Such
interactions could affect only the unobservable part of the different
circles (i.e., the sets $\mathcal{S}$, $\mathcal{A}$, and
$\mathcal{R}a$).
In this case we can expect the probability
transition matrix would have dominant diagonal blocks and a few off
diagonal terms accounting for the interaction. The synthesis of the optimal control
policy could be done applying the technique suggested by~\citeauthor{Hak18}~\cite{Hak18}.
A second possibility is to consider the abstraction technique suggested~\citeauthor{nasir2020epidemics}~\cite{nasir2020epidemics}.
In this case each individual would be considered as the representative of a few tens of people. A key issue
would be in this case to produce a realistic analytical evaluation of the error made in probability.
}
Fourth, carry out an accurate tuning of the probabilities, given the
historical series that are now available. However, due to the
heterogeneity of the aggregated data and the partial coverage with
respect to the model state, this is a non trivial task.
%
From a technological point of view, we aim at a tighter integration of
our proof of concept tool with \texttt{PRISM} to avoid the generation and
parsing of large files making the analysis of the full model tractable
for meaningful population sizes.
Moreover, we want to investigate the use of abstraction techniques
to improve performance \dan{and, thus, including in the picture the
vaccinated population, which may be studied to establish the
interventions to reach, e.g., herd immunity.}
Finally, we want to investigate the use of POMDP and the automatic
synthesis of policies achieving a given PLTL property.

%
\noindent
{\small \textbf{Acknowledgements. } This work is partially funded with project
40900028, Bando interno 2020 Università di Trento "Covid 19", lead by Marco Roveri.}

\bibliography{mainAutomatica}       


\appendix
\clearpage
\section{SUPPLEMENTAL MATERIAL - Complete proof of Theorem~\ref{teo:trasProb}}
\label{sec:appA}
By applying the theorem of total probability in the conditioned case
to~\eqref{eq:transProbReDef}, we have:
\[
  \begin{aligned}
    & \Prc{\vec{V}_{k+1}=\vec{v}'}{\vec{V}_k} = \Prc{l(\vec{\Delta}) \models B}{\vec{V}_k} = \\
    & \sum_{\delta_1 \in D} \!\!\Prc{l(\vec{\Delta}) \models B}{\vec{V}_k
      \!\wedge\! (\Delta_1=\delta_1)}\Prc{\Delta_1=\delta_1
    }{\vec{V}_k}\! .
  \end{aligned}
\]
From the first balance equation~\eqref{eq:balanceTests}, we observe
that
\[
  l(\vec{\Delta}) \models B \implies
  (\Delta_1=\delta_1) \models B_1 \implies
  \delta_1 = - \Delta_S.
\]
When $\delta_1 \neq -\Delta_S$, it implies that
$$
 \Prc{l(\vec{\Delta}) \models B}{\vec{V}_k \wedge (\Delta_1=\delta_1)} = 0,
$$
which makes all the terms in the summation vanish but the term
$\Delta_1=-\Delta_S$. Applying the above introduced shorthand $l_1$,
the transition probability is equal to
\begin{equation}
  \Prc{l(\vec{\Delta}) \models B}{\vec{V}_k \wedge l_1}
  \cdot
  \Prc{l_1}{\vec{V}_k}.
\end{equation}

We can then iterate the application of the total probability theorem on
the first term of \eqref{eq:l1} as follows
\[
  \begin{aligned}
    & \sum_{(\delta_2,\delta_3,\delta_9) \in D^3}
    \!\!\!\!\!\!\!\!\!\!\Prc{l(\vec{\Delta}) \models B}{\vec{V}_k \wedge l_1 \wedge l_5}
     \cdot \Prc{l_5}{\vec{V}_k \wedge l_1 }.
  \end{aligned}
\]
We can again observe that the triple summation is only apparent, since
for the terms that do not satisfy $B_2$, that is,
$(\Delta_1=\delta_1, l_5) \nvDash B_2 $, it holds that
\[
  \Prc{l(\vec{\Delta}) \models B}{\vec{V}_k \wedge l_1 \wedge
    l_5} = 0.
\]
As in the previous case, considering the second balance equation
\eqref{eq:balanceTests}, we have introduced the shorthand $l_2$ which
simplifies the notation for the above term to:
\[
\begin{aligned}
& \Prc{l(\vec{\Delta}) \models B}{\vec{V}_k \wedge l_1 \wedge
    l_5} = \\
& \sum_{\delta_2=0}^{-\Delta_S-\Delta_A-\Delta_{R_a}}
  \!\!\!\!\!\!\!\!\!\!\!\!\!\!\Prc{l(\vec{\Delta}) \models B}{\vec{V}_k \wedge l_1  \wedge l_2 }
   \Prc{l_2}{\vec{V}_k \wedge l_1 }.
\end{aligned}
\]
We can apply once again the theorem of total probability and enforce
the validity of $B_3$ in shorthand $l_3$:
\[
\begin{aligned}
  &\Prc{l(\vec{\Delta}) \models B}{\vec{V}_k \wedge l_1 \wedge l_2}\\
  &= \sum_{\delta_4=0}^{\delta_2 -
    \Delta_I}\sum_{\delta_5=0}^{\delta_2 - \Delta_I-\delta_4}
  \sum_{\delta_6=0}^{\delta_2 - \Delta_I-\delta_4-\delta_5} \Prc{l_3}{\vec{V}_k \wedge l_1 \wedge l_2}\cdot\\
  &\,\,\,\, \cdot \Prc{l(\vec{\Delta}) \models B}{\vec{V}_k \wedge l_1
    \wedge l_2 \wedge l_3}.
\end{aligned}
\]
Our last iteration of the theorem of total probability that enforces
the validity of $B_4$, $B_5$ and $B_6$ by means of the shorthand
$l_4$, leads to:
\[
\begin{aligned}
  &\Prc{l(\vec{\Delta}) \models B}{\vec{V}_k \wedge l_1 \wedge l_2 \wedge l_3}  =\\
  &= \sum_{(\delta_7, \delta_8, \delta_{10}, \delta_{11}) \in D^4}  \Prc{l_4}{\vec{V}_k \wedge l_1 \wedge l_2 \wedge l_3 } \cdot \\
  &\,\,\,\, \cdot \Prc{l(\vec{\Delta}) \models B}{\vec{V}_k \wedge l_1
    \wedge l_2 \wedge l_3 \wedge l_4},
\end{aligned}
\]
It is easy to observe that
$\vec{V}_k \wedge l_1 \wedge l_2 \wedge l_3 \wedge l_4$ is a certain
event, thus
\[
\Prc{l(\vec{\Delta}) \models B}{\vec{V}_k = \vec{v} \wedge l_1 \wedge l_2  \wedge l_3 \wedge l_4 } =1 ,
\]
which leads us to the final simplification and yields the theorem.

\section{SUPPLEMENTAL MATERIAL Simplified Markov Model: The Case of Untested Subjects}
\label{sec:markov_untestes}
In this section, we focus on the case in which subjects are not tested
(and hence are not quarantined), which is usually the situation at the
beginning of a disease as \covid{} where no clear medical test were
known or developed. The possible states of a subject are the ones
depicted in Figure~\ref{fig:stateTest} where
$\Delta_3 = \Delta_9 = \Delta_{10} = \Delta_{11} = 0$ and where
$|Q_k| = |R_{a,k}| = 0$, $\forall k$.  Hence, the state of the Markov
chain will be only associated to the 6-tuple given by
$(S_k, A_k, I_k, R_k, O_k, D_k)$, while the
constraints~\eqref{eq:domain_invariantTested}, the balance
equations~\eqref{eq:balanceTests} and the flow
constraints~\eqref{eq:mr1Tests} are still valid under the previously
reported assumption. As a consequence, the dynamic of the state vector
\[
\vec{V}_k =
  \begin{bmatrix}
     S_k & A_k & I_k & R_k & O_k & D_k
  \end{bmatrix}^T ,
\]
will obey the transition probability rules in~\eqref{eq:transProbDef},
where of course
\[
  \vec{\Delta v} = \vec{v}'-\vec{v} =
  \begin{bmatrix}
     \Delta_S & \Delta_A & \Delta_I & \Delta_R & \Delta_O & \Delta_D
  \end{bmatrix}^T.
\]

With the same steps of the previous case and besides the variable
assignment $l_1(\cdot)$, we introduce the shorthand notations for the
variable assignments:

\begin{itemize}

\item $l_2$, defined as a function depending of $\delta_2$ and
  $\delta_3$, with the variable $\delta_3$ obtained via equation
  $B_2$,
  that is \\
  $l_2$:
  $(\Delta_2=\delta_2, \Delta_3=-\Delta_S-\Delta_A - \delta_2)$;

\item $l_3$, defined as a function depending of
  $\delta_4,\delta_5,\delta_6$, that is $l_3(\delta_4,\delta_5)$ and
  with $\delta_6$ obtained via equation $B_3$,\\
  $l_3$:
  $(\Delta_4=\delta_4,\Delta_5=\delta_5,
  \Delta_6=\delta_2-\Delta_I-\delta_4-\delta_5)$;

\item $l_4$, defined as a function depending of
  $\delta_2,\delta_4,\delta_5$, with the relations obtained from $B_4$
  and $B_5$, \\
  $l_4$:
  $(\Delta_7 = \Delta_D + \Delta_I - \delta_2 + \delta_4 + \delta_5,
  \Delta_8 = \delta_2-\delta_4- \Delta_D-\Delta_I-\Delta_O)$.

\end{itemize}

Proceeding as in Section~\ref{sec:markov_tested}, we can state the
following:

\begin{theorem}
The transition probability~\eqref{eq:transProbReDef} is re-written as
  \begin{equation}\label{eq:transProbFinal}
  \scriptsize
    \begin{array}{l}
       \Prc{\vec{\Delta} \models B}{\vec{V}_k = \vec{v}}  =
       \Prc{l_1}{\vec{V}_k=\vec{v}} \!\!\!\!\!\!\displaystyle\sum_{\delta_2=0}^{-\Delta_S-\Delta_A}
       \!\!\!\!\Prc{l_2}{\vec{V}_k \wedge l_1} \\
       \displaystyle\sum_{\delta_4=0}^{\delta_2 - \Delta_I}
       \!\sum_{\delta_5=0}^{\delta_2 - \Delta_I-\delta_4}
       \!\!\!\!\!\!\!\!  \Prc{l_3}{\vec{V}_k \wedge l_1 \wedge l_2}
       \!\cdot\! \Prc{l_4}{\vec{V}_k \wedge l_1 \wedge l_2 \wedge l_3}.
    \end{array}
  \end{equation}
\end{theorem}

The proof is a simplified version of Theorem~\ref{teo:trasProb}.  We
can now move to the computation of the four different probabilities
that appear in~\eqref{eq:transProbFinal}.

\subsection{Probability that a susceptible subject becomes infectious}

The probability of $\Prc{g_k}{\vec{V}_k}$ in this simplified case
follows exactly the same lines of Section~\ref{subsec:ProbInfected},
where, obviously, \eqref{eq:ProbInfectedTested} turns to
\[
\begin{aligned}
  \Prc{g_k}{\vec{V}_k}_M = 1 - \left( 1 - \omega
    \frac{A_k}{N-D_k-I_k-O_k} \right)^{M}.
\end{aligned}
\]

\subsection{Probability of state change for a number of subjects}
\label{sec:changeSimpleStates}

The probabilities of state change in~\eqref{eq:transProbFinal} can be
computed considering as reference Figure~\ref{fig:stateTest}, where
$\Delta_3 = \Delta_9 = \Delta_{10} = \Delta_{11} = 0$ and where
$|Q_k| = |R_{a,k}| = 0$, $\forall k$.

{\bf Transition from State $S$.}  Nothing changes with respect to
Section~\ref{sec:markov_tested}. Hence, the term
$\Prc{l_1}{\vec{V}_k}$ is the same as in
equation~\eqref{eq:FirstCombProb}.

{\bf Transition from state $A$:} Since the probability of the event
``exactly $\delta_3$ asymptomatic infected subjects recover at step
$k$ and $\delta_2$ subjects become symptomatic'' is given by
Theorem~\ref{th:RhoTheorem}, for the assignment $l_2$, we have
\begin{equation}
  \Prc{l_2}{\vec{V}_k \wedge l_1} = \rho(\delta_2, -\Delta_S-\Delta_A-\delta_2, \vec{V}_k).
  \label{eq:SecondCombProb}
\end{equation}

{\bf Transition from state $I$:} With respect to the case considered
in Section~\ref{sec:markov_tested}, the only difference is given by
the constrained value of $\delta_6$, i.e.
\begin{equation}
 \Prc{l_3}{\vec{V}_k \wedge l_1 \wedge l_2} = \phi(\delta_4, \delta_5, \delta_2-\delta_4-\delta_5-\Delta_I, \vec{V}_k).
 \label{eq:ThirdCombProb}
\end{equation}

{\bf Transition from State $O$:} This case is identical to the
previous case of Section~\ref{sec:markov_tested}, so we have
\begin{equation}
  \label{eq:FourthCombProb}
  \begin{aligned}
    &\Prc{l_4}{\vec{V}_k \wedge l_1 \wedge l_2 \wedge l_3}  \\
    &= \zeta ( \delta_D + \delta_I \!-\! \delta_2 + \delta_4 +
    \delta_5,
    \delta_2\!-\!\delta_4\!-\!\Delta_D\!-\!\Delta_I\!-\!\Delta_O,
    \vec{V}_k) .
  \end{aligned}
\end{equation}

The computation of the transition probabilities is thus obtained by plugging~\eqref{eq:FirstCombProb},~\eqref{eq:SecondCombProb},~\eqref{eq:ThirdCombProb} and~\eqref{eq:FourthCombProb} into ~\eqref{eq:transProbFinal}.

\end{document}